\documentclass[aps,prd,twocolumn,superscriptaddress,preprintnumbers,floatfix,nofootinbib,notitlepage,showkeys,showpacs]{revtex4-1}

\usepackage[utf8]{inputenc}
\usepackage{cancel}

\usepackage{graphicx}
\usepackage{hyperref}
\usepackage{latexsym}
\usepackage{amsmath}
\usepackage{amssymb}
\usepackage{bbm}

\usepackage{xcolor}
\definecolor{cites}{RGB}{0,180,0}
\definecolor{links}{RGB}{200,0,0}
\hypersetup{colorlinks=true,citecolor=cites,linkcolor=links,urlcolor=blue}

\usepackage[normalem]{ulem}
\usepackage{pdfsync}
\usepackage{epsfig}
\usepackage{epstopdf}
\usepackage{subfigure}
\usepackage{color}
\usepackage{comment}
\usepackage{slashed}
\usepackage{placeins}
\usepackage{cleveref}

\newcommand{\be}{\begin{equation}} 
\newcommand{\ee}{\end{equation}}
\newcommand{\bea}{\begin{eqnarray}} 
\newcommand{\eea}{\end{eqnarray}}

\newcommand{\bmp}{\noindent\begin{minipage}{16cm}}
\newcommand{\emp}{\end{minipage}\vskip 7mm} 
\def\lsim{\mathrel{\raise.3ex\hbox{$<$\kern-.75em\lower1ex\hbox{$\sim$}}}}
\def\gsim{\mathrel{\raise.3ex\hbox{$>$\kern-.75em\lower1ex\hbox{$\sim$}}}}
\newcommand{\intron}[1]{}

\newcommand{\MSb}{\overline{\textrm{MS}}}

\newcommand{\eq}[1]{Eq.~(\ref{#1})}

\newcommand{\A}{{a}}
\newcommand{\gGF}{g_{\rm GF}^2}

\newcommand{\krsout}[1]{}

\begin{document}

\title{Infrared fixed point of SU(2) gauge theory with six flavors}

\author{Viljami Leino}
\email{viljami.leino@helsinki.fi}
\affiliation{Department of Physics, University of Helsinki \\
                      P.O.~Box 64, FI-00014, Helsinki, Finland}
\affiliation{Helsinki Institute of Physics, \\
                      P.O.~Box 64, FI-00014, Helsinki, Finland}

\author{Kari Rummukainen}
\email{kari.rummukainen@helsinki.fi}
\affiliation{Department of Physics, University of Helsinki \\
                      P.O.~Box 64, FI-00014, Helsinki, Finland}
\affiliation{Helsinki Institute of Physics, \\
                      P.O.~Box 64, FI-00014, Helsinki, Finland}

\author{Joni Suorsa}
\email{joni.suorsa@helsinki.fi}
\affiliation{Department of Physics, University of Helsinki \\
                      P.O.~Box 64, FI-00014, Helsinki, Finland}
\affiliation{Helsinki Institute of Physics, \\
                      P.O.~Box 64, FI-00014, Helsinki, Finland}

\author{Kimmo Tuominen}
\email{kimmo.i.tuominen@helsinki.fi}
\affiliation{Department of Physics, University of Helsinki \\
            P.O.~Box 64, FI-00014, Helsinki, Finland}
\affiliation{Helsinki Institute of Physics, \\
            P.O.~Box 64, FI-00014, Helsinki, Finland}

\author{Sara Tähtinen}
\email{sara.tahtinen@helsinki.fi}
\affiliation{Department of Physics, University of Helsinki \\
            P.O.~Box 64, FI-00014, Helsinki, Finland}
\affiliation{Helsinki Institute of Physics, \\
            P.O.~Box 64, FI-00014, Helsinki, Finland}

\begin{abstract}
We compute the running of the coupling in SU(2) gauge theory with six fermions in the fundamental representation of the gauge group.
We find a strong evidence that this theory has an infrared stable fixed point at strong coupling
and measure also the anomalous dimension of the fermion mass operator at the fixed point.
This theory therefore likely lies close to the
boundary of the conformal window and will display novel infrared dynamics
if coupled with the electroweak sector of the Standard Model.
\end{abstract}

\preprint{HIP-2017-03/TH}

%
\maketitle

\section{Introduction}\label{sec:introduction}
Determination of the vacuum phase of an SU($N$) gauge theory as a function
of the number of massless flavors of Dirac fermions, $N_f$, and their
representations presents a challenge for our basic understanding of gauge
theory dynamics at strong coupling.
A lot of effort in the field of lattice gauge theory has been devoted to address the existence
and properties of infrared fixed point (IRFP),
which appears when $N_f$ is between a critical lower limit $N_f^{\rm{crit}}$ and the loss of asymptotic freedom.
The bounds depend on $N$ and the fermion representation.
For recent reviews see~\cite{Pica:2017gcb,Nogradi:2016qek,DeGrand:2015zxa}.
A much studied benchmark case is SU(2) gauge theory with two Dirac
fermions in the adjoint
representation~\cite{Hietanen:2008mr,Hietanen:2009az,
  DelDebbio:2008zf,Catterall:2008qk,
  Bursa:2009we,DelDebbio:2009fd,DelDebbio:2010hx,
  DelDebbio:2010hu,Bursa:2011ru,DeGrand:2011qd, Patella:2012da,
  Giedt:2012rj,DelDebbio:2015byq,
  Rantaharju:2015yva,Rantaharju:2015cne},
where the results indicate the existence of an IRFP.

In SU(2) gauge theory with fermions in the fundamental representation
the precise dependence on $N_f$ remains uncertain, despite  a large number of
recent studies on the lattice~\cite{Ohki:2010sr,Bursa:2010xn,Karavirta:2011zg,Hayakawa:2013maa,Appelquist:2013pqa,Leino:2017lpc}.
The upper edge of the conformal window is robust:
the asymptotic freedom is lost at $N_f=11$, where the 1-loop $\beta$-function
coefficient changes sign.
Just below the upper edge, at 10 flavors the theory has a perturbatively stable Banks-Zaks type infrared fixed point \cite{Banks:1981nn}, which has also been observed on the lattice~\cite{Karavirta:2011zg}.  Recently, simulations of the 8 flavor theory have also shown the existence of a fixed point~\cite{Leino:2017lpc}.
On the other hand, the theory with $N_f=2$ is well below the conformal window and breaks the chiral symmetry according to the expected pattern,
and the theory with $N_f=4$ is expected to fall within this category as well~\cite{Karavirta:2011zg}.
However, for $N_f=6$ the previous results remain so far inconclusive~\cite{Bursa:2010xn,Karavirta:2011zg,Hayakawa:2013maa,Appelquist:2013pqa}.

In perturbation theory the $\beta$-function is known up to 5-loop order in the $\MSb$ scheme \cite{Herzog:2017ohr}.
In SU(2) gauge theory with $N_f=6$ fundamental representation fermions the $\beta$-function has a non-trivial zero (i.e an IRFP) up to 4-loop order. 
In the 5-loop expansion of the $\beta$-function the IRFP vanishes. 
Similar behavior has been observed in SU(3) with $N_f=12$~\cite{Fodor:2016zil}.
However, the SU(2) 5-loop $\beta$-function shows peculiar behaviour as $N_f$ is varied:
it predicts an IRFP in two disconnected domains, at $3.0 \lsim N_f \lsim 5.8$ and 
$8.6 \lsim N_f < 11$.  $N_f=6$ lies between these ranges.  This kind of behaviour is clearly
unphysical, and shows that perturbation theory cannot be quantitatively relied upon when the fixed point appears at strong coupling.

In this article we give strong evidence that
the six flavor theory indeed has an IRFP at strong coupling.
The result is based on a thorough state-of-the-art measurements of the running coupling
and the anomalous dimension of the fermion mass operator.  We use the
HEX smeared Wilson-clover fermion lattice action and measure the coupling using the Yang-Mills gradient flow~\cite{Narayanan:2006rf,Luscher:2009eq} in conjunction with the finite volume step scaling function with Dirichlet (``Schrödinger functional'') boundary conditions \cite{Ramos:2015dla}. 
The value of the coupling at IRFP, $g^2_\ast$, is scheme dependent and hence depends on the gradient flow time.  With our benchmark scheme we find $g_\ast^2 = 14.5(4)_{-1.2}^{+0.4}$ 
with statistical and systematic errors.

We also measure two scheme independent quantities at the IRFP: the mass anomalous dimension
$\gamma_m^\ast$ and the leading irrelevant critical exponent $\gamma_g^\ast$,
which gives the slope of the $\beta$-function at IRFP.  The mass anomalous dimension is measured using two different methods: the mass step scaling method~\cite{Capitani:1998mq} and the Dirac operator spectral density method \cite{Patella:2011jr}.  At the fixed point we observe $\gamma_m^\ast = 0.283(2)^{+0.01}_{-0.01}$.  The slope of the $\beta$-function is directly
measurable from the step scaling function of the coupling, obtaining
$\gamma_g^\ast=0.648(97)_{-0.1}^{+0.16}$. 
In contrast to the value of the fixed point coupling, we observe that $\gamma_g^\ast$ remains independent of the gradient flow time, in accord with the scheme independence of this quantity.

This paper is structured as follows: 
In section~\ref{sec:impl} we define the model and outline the simulation methods.
The numerical results are presented for running coupling, leading irrelevant exponent, and mass anomalous dimension, in 
sections~\ref{sec:evolution},~\ref{sec:gammag}, and~\ref{sec:anomalous} respectively. 
In section~\ref{sec:conclude} we present our conclusions.

\vspace{2mm}
\section{Lattice implementation}\label{sec:impl}
The model we use and the algorithmic details we apply are described
in detail in~\cite{Rantaharju:2015yva,Leino:2017lpc}, and our discussion
here will be brief so that we can then focus on the results we obtain in
the case of $N_f=6$.
The model is defined by the lattice action
\begin{displaymath}
	S = (1-c_g)S_G(U) + c_g S_G(V) + S_F(V) + c_{\rm SW} \delta S_{SW}(V) \,.
\end{displaymath}%
The smeared gauge link $V$ is defined by hypercubic truncated stout smearing (HEX smearing)~\cite{Capitani:2006ni},
and we mix smeared, $S_G(V)$, and unsmeared, $S_G(U)$, Wilson gauge actions with
mixing parameter $c_g=0.5$.
This partial smearing allows us to reach significantly larger couplings
by avoiding the unphysical bulk phase transition in the region of interest of
the parameter space~\cite{DeGrand:2011vp}.
We use clover Wilson fermion action with the Sheikholeslami-Wohlert coefficient set to
tree level value of unity, $c_{\rm SW}= 1$, which is the standard choice for smeared clover fermions.  We have verified that this value is very close to the true non-perturbatively fixed $c_{\rm SW}$ coefficient, canceling most of the $O(a)$ errors.

On a lattice of size~$L^4$ we use Dirichlet boundary conditions
at the temporal boundaries $x_0=0$, $L$ by setting
the gauge link matrices $U=V=1$ and the fermion fields to zero.
The spatial boundaries are periodic.
These boundary conditions enable simulations at vanishing fermion mass, and
allow the mass anomalous dimension to be measured using the same configurations
as for the running coupling.

We run our simulations using the hybrid Monte Carlo algorithm
with 2nd order Omelyan integrator~\cite{Omelyan:2003:SAI,Takaishi:2005tz}
and chronological initial values for the fermion matrix inversions~\cite{Brower:1995vx}.
We tune the step length to have an acceptance rate larger than 85\%.
We run the simulations with bare couplings varying within the range
\begin{equation}
\beta_L \equiv 4/g_0^2  \in[0.5,8]
\end{equation}
and tune the hopping parameter $\kappa_c(\beta_L)$
so that the absolute value of the PCAC fermion mass~\cite{Luscher:1996vw}
is less than $10^{-5}$ at lattices of size $24^4$.
The same critical hopping parameter values are used for all the lattice sizes,
and for each $\beta_L$ (and corresponding $\kappa_c(\beta_L)$).
The critical hopping parameters are available in the Table~\ref{tab:kappa_sup1}.
We use lattices of size $L=$8, 10, 12, 16, 18, 20, 24, and 30
chosen to allow step scaling with either $s=2$ or $s=3/2$.
For our analysis we choose $s=3/2$ as it includes more pairs within the larger lattices.
We generate $(5-100)\cdot 10^{3}$ trajectories for each combination of $\beta_L$ and $L$.
For the exact number of trajectories used, see Table~\ref{tab:trajnum_sup2}.

To define the running coupling,
we apply the Yang-Mills gradient flow method~\cite{Narayanan:2006rf,Luscher:2009eq,Ramos:2015dla}.
This method defines a flow that smooths the gauge fields and removes UV divergences
and automatically renormalizes gauge invariant objects~\cite{Luscher:2011bx}.
The method is set up by introducing a fictitious flow time $t$
and studying the evolution of the flow gauge field $B_\mu(x,t)$ according to the flow equation
\begin{equation}
  \partial_t B_{\mu} = D_{\nu} G_{\nu\mu} \,,\;
  \label{eq:gradflow}
\end{equation}%
where $G_{\mu\nu}(x;t)$ is the field strength of the flow field $B_{\mu}$ and
$D_\mu=\partial_\mu+[B_\mu,\,\cdot\,]$.
The initial condition is $B_{\mu}(x;t=0) = A_\mu(x)$,
where $A_\mu$ is the original continuum gauge field.
In the lattice formulation the (unsmeared) lattice link variable $U$
replaces the continuum flow field,
which we then evolve using either the tree-level
improved L\"uscher-Weisz pure gauge action (LW)~\cite{Luscher:1984xn}
or the Wilson plaquette gauge action (W).  In the continuum limit these 
should yield identical results, providing a check of the reliability of the limit.

The coupling at scale $\mu=1/\sqrt{8t}$~\cite{Luscher:2010iy} is defined via energy measurement as
\begin{align}
	\label{eq:g2gf}
	\gGF(\mu) &= \mathcal{N}^{-1}t^2 \langle E(t+\tau_0 a^2) \rangle\vert_{x_0=L/2\,,\,t=1/8\mu^2}\,,
\end{align}%
where $a$ is the lattice spacing.
The shift parameter $\tau_0$ is introduced to reduce the $\mathcal{O}(a^2)$
discretization effects caused by the flow~\cite{Cheng:2014jba}
and can be numerically estimated during the analysis.
The normalization factor $\mathcal{N}$ for the chosen boundary conditions has
been calculated in~\cite{Fritzsch:2013je}
to match the $\MSb$ coupling in the tree level.
As the translation symmetry is broken by the chosen boundary conditions,
the coupling $\gGF$ is measured only on the central time slice $x_0=L/2$.
To quantify the effects of different discretizations,
we measure the energy density $E(t)$ using both symmetric clover and simple plaquette discretizations of the flow.

In order to limit the scale into a regime
where both lattice artifacts and finite volume effects are minimized,
we relate the lattice and the renormalization scales by defining a dimensionless parameter $c_t$ such that
$\mu^{-1} = c_tL = \sqrt{8t}$ as described in~\cite{Fodor:2012td,Fritzsch:2013je}.
The chosen boundary conditions have a reasonably small cutoff effects and statistical variance
within the range of $c_t=0.3-0.5$~\cite{Fritzsch:2013je}.
Value of this parameter defines the renormalization scheme.

\vspace{2mm}
\section{Evolution of the coupling}
\label{sec:evolution}
Our ``benchmark'' set of results presented here are obtained with gradient flow~\eq{eq:gradflow} evolved with L\"uscher-Weisz action (LW),
clover definition of energy density~\eq{eq:g2gf} and $c_t=0.3$.
In order to estimate systematic errors, we vary discretizations of the flow
and the observable and the parameter values of the flow.
The raw data is available in
\cref{tab:g2c03wt_sup3,tab:g2c03_sup4,tab:g2c035_sup5,tab:g2c04_sup6,tab:g2c045_sup7,tab:g2c03W_sup8,tab:g2c03P_sup9,tab:g2c03PW_sup10}.

\begin{figure}[t]
  \includegraphics[width=8.6cm]{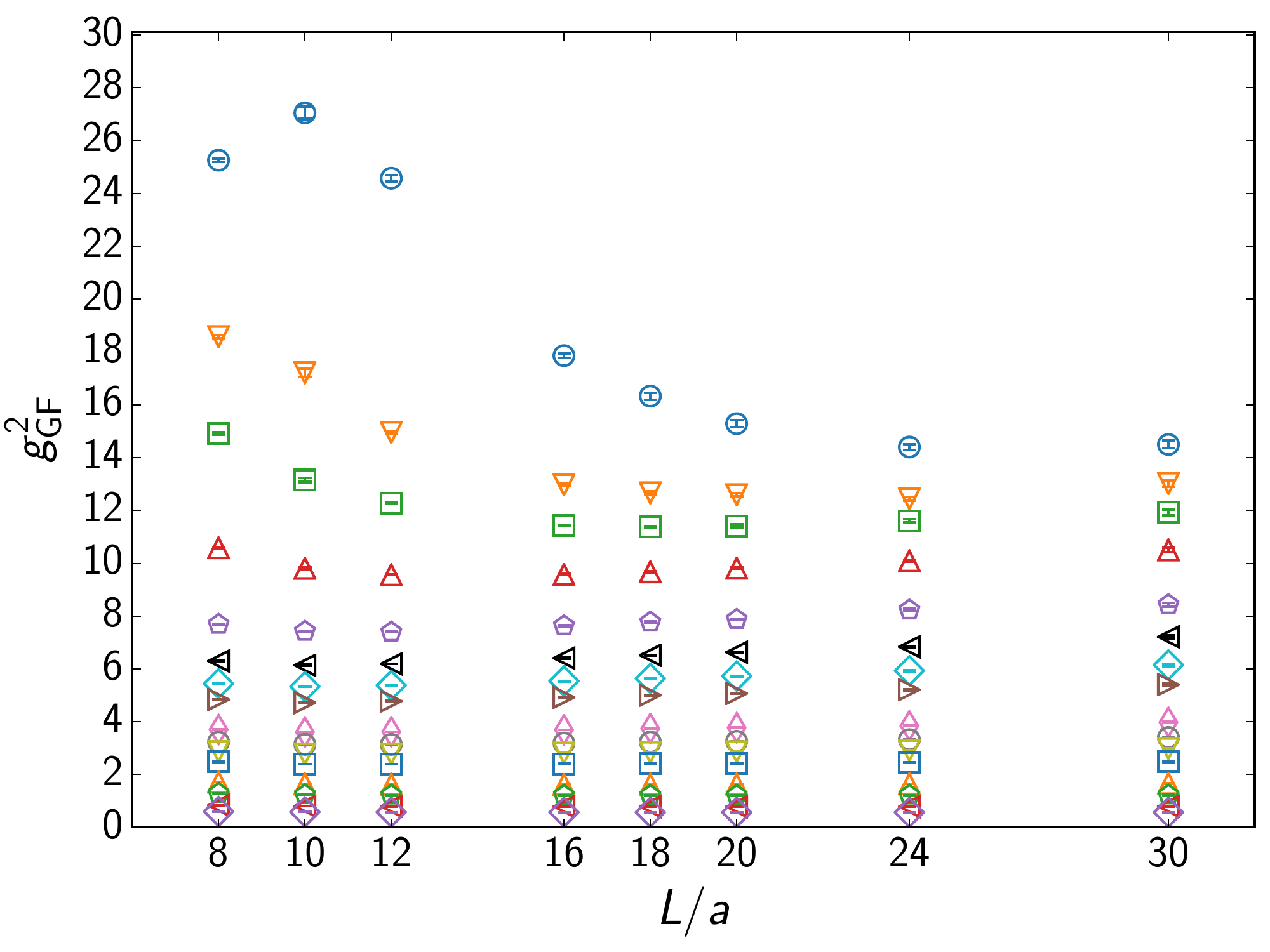}
  \caption[b]{The gradient flow coupling~\eq{eq:g2gf}
    measured at each $\beta_L$ and $L/\A$ using the benchmark set
    of parameters (LW flow action, clover definition of field strength, $c_t=0.3$).}
  \label{fig:g2_lat_meas}
\end{figure}

The measured couplings with the aforementioned parameters are shown in the top panel of Fig.~\ref{fig:g2_lat_meas}.
It is clear from the figure that the finite volume effects become substantial on smaller lattices as the
coupling grows larger.

\begin{figure}[t]
  \includegraphics[width=8.6cm]{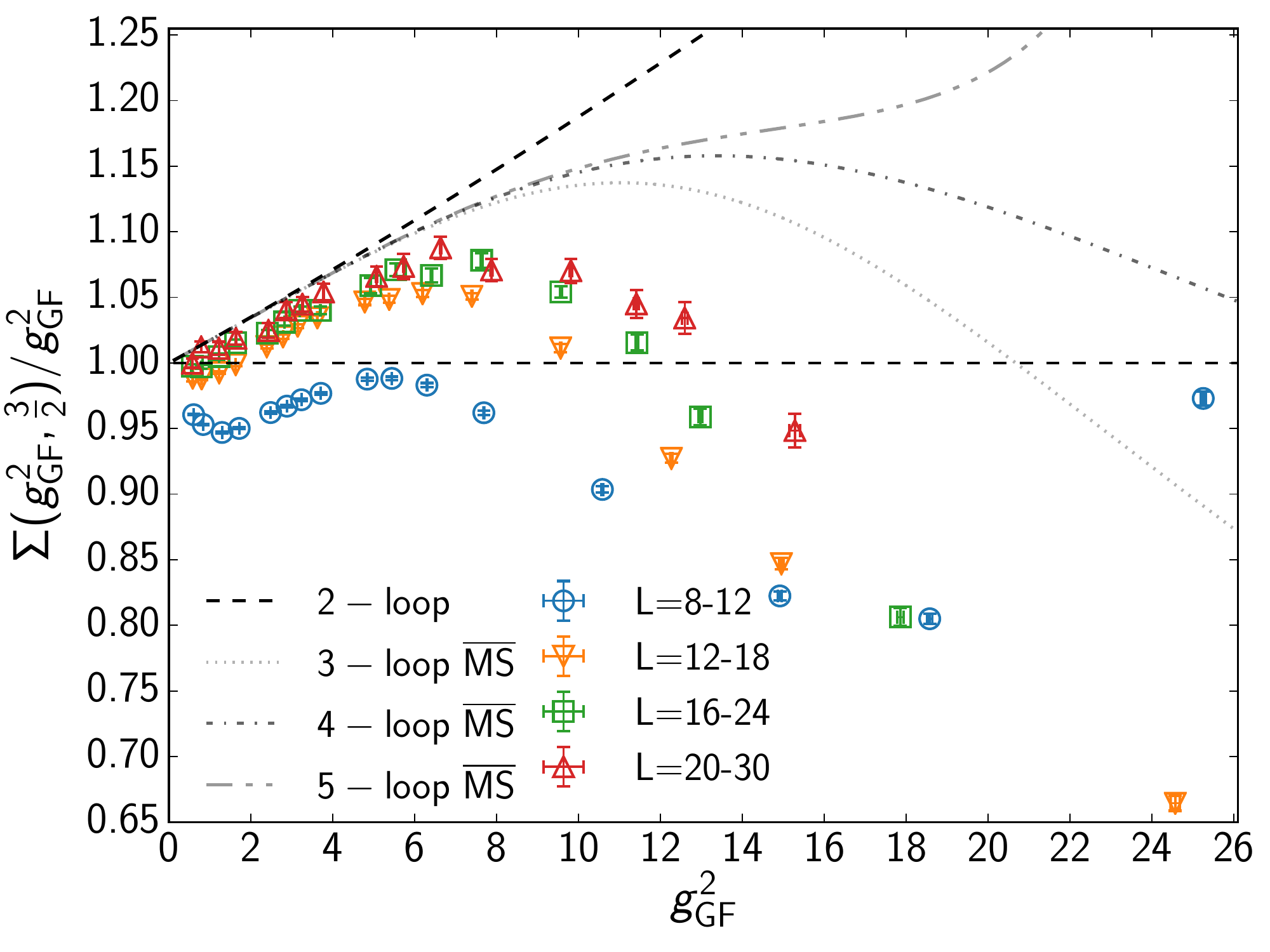}
  \caption[b]{The lattice step scaling function~\eq{eq:lat_step_raw} obtained from the
    data in Fig.~\ref{fig:g2_lat_meas}.
  }
  \label{fig:g2_lat_step}
\end{figure}%

To quantify the running of the coupling we use the finite volume step
scaling function~\cite{Luscher:1993gh}:
\begin{equation} \label{eq:lat_step_raw}
    \Sigma(u,L/\A,s) = \left . \gGF(g_0,sL/\A) \right|_{\gGF(g_0,L/\A)=u}\,,
\end{equation}%
which describes the change of the measured coupling
when the linear size of the system is increased from $L$ to $sL$.
The step scaling obtained from the data in Fig.~\ref{fig:g2_lat_meas} is shown
in Fig.~\ref{fig:g2_lat_step}, using a scaling factor $s=3/2$, that is
pairs of lattices with $L/a =
8-12$, $12-18$, $16-24$ and $20-30$.
At the smallest volume pair $L/a = 8-12$ the step scaling deviates significantly from the others,
and will be excluded from the continuum analysis.
Rest of the volume pairs are observed to follow the scheme independent 2-loop curve in the weak coupling region up to $\gGF\sim6$
after which the the measured step scaling function deviates towards an IRFP around $\gGF\sim14$.
While the higher loop $\MSb$ results are scheme dependent and cannot be directly compared with our result,
we show them for comparison. 

\begin{table}
\begin{ruledtabular}
\begin{tabular}{cccccccc}
 $c_t$ & L=8a & L=12a & L=16a & L=18a & L=20a & L=24a & L=30a \\
\hline
0.3  & 1.3   & 0.5 & 1.0 & 1.1 & 0.9 & 1.5 & 1.5 \\
0.35 & 12    & 0.4 & 0.6 & 0.8 & 0.9 & 1.3 & 1.4 \\
0.4  & 9.3   & 0.7 & 0.4 & 0.7 & 1.1 & 1.2 & 1.6 \\
0.45 & 8.6   & 1.1 & 0.5 & 1.1 & 1.4 & 1.1 & 1.9 \\
\end{tabular}
\end{ruledtabular}
\caption{\label{tab:chi2}
$\chi^2/$d.o.f of the fit \eqref{eq:betafitfun} at different $L/a$ and $c_t$.
}
\end{table}%

For the continuum limit $\sigma(\gGF)$ of the step scaling function we use the
extrapolating function
\begin{align} \label{eq:lat_step_cont}
  \Sigma(\gGF,L/\A)  &= \sigma(\gGF) + c(\gGF) (\A/L)^2\,.
\end{align}%
At weak coupling the cutoff effects are regulated by the proximity of
the ultraviolet fixed point, and the lowest order discretization
effects of the Wilson-clover action are expected to be of order
$\mathcal{O}(a^2)$, motivating the use of Eq.~\eqref{eq:lat_step_cont}.  

As will be seen below, at small couplings the $O(\A^2/L^2)$ extrapolation works quite well.
However, at large couplings the range of volumes available to us and the accuracy of measurements
are not sufficient to verify this.  Using staggered fermions and much larger volumes, it has been observed that in SU(3) theory with $N_f=12$ fundamental fermions including $O(\A^4/L^4)$ effects can affect the continuum limit at a few 10\% level \cite{Fodor:2017gtj}.

At large coupling, so long as the coupling remains below
the possible IRFP, the continuum limit is ultimately
reached at the UV fixed point.  However, due to the smallness of the
$\beta$-function this would require astronomically large scale
hierarchy between the lattice size $L$ and lattice spacing $a$ and
hence is impossible to observe in simulations.  Nevertheless, if the
anomalous exponents of the fields remain small near the infrared fixed point, one can assume
that the power counting of operators is applicable and the cutoff effects (dominated
by dimension 6 operators) decrease with a power of the lattice spacing $a$.  
The naive $a^2$ behaviour may be modified by anomalous exponents, though.
In section \ref{sec:anomalous} we observe that the mass anomalous dimension
at the IRFP remains relatively small, $\gamma_m \approx 0.28$, suggesting that the
$a^2$ behaviour in Eq.~\eqref{eq:lat_step_cont} may also receive only minor corrections.
The available range in our data does not allow us to numerically determine 
differences from Eq.~\eqref{eq:lat_step_cont} and hence we use it at all couplings.
Near the IRFP we indeed observe that the continuum limit becomes somewhat less robust, 
which is taken into account in our systematic error estimation.

In order to determine the continuum limit of the step scaling function we need 
the measurements of the step scaling at constant value of the coupling. However,
in practice the simulations are carried out at a fixed set of bare lattice couplings
which do not correspond to same value of $\gGF$ when $a/L$ is varied.
Hence, we adopt the customary interpolation procedure of the measured couplings to intermediate
couplings.  We do this using a polynomial fit\footnote{A rational interpolating function
is another choice used in the literature \cite{Karavirta:2011zg,Leino:2017lpc}.   However, in our
case this did not offer any improvement.}
\begin{equation} \label{eq:betafitfun}
    \gGF(g_0,\A/L) = g_0^2(1+\sum_{i=1}^m  a_i g_0^{2i}) \,, 
\end{equation}%
where we use $m=10$ for lattices smaller than $L=16$ and $m=9$ for the larger lattices.
With this choice we obtain the $\chi^2/$d.o.f's reported in Table~\ref{tab:chi2} for each used $c_t$
and in Table~\ref{tab:combchi} for each used discretization. 
We study the robustness of the fits by repeating the analysis with $m$ decreased by one.
While this choice increases the $\chi^2$/d.o.f., the results stay compatible with those 
obtained with larger $m$.

\begin{figure}[t]
  \includegraphics[width=8.6cm]{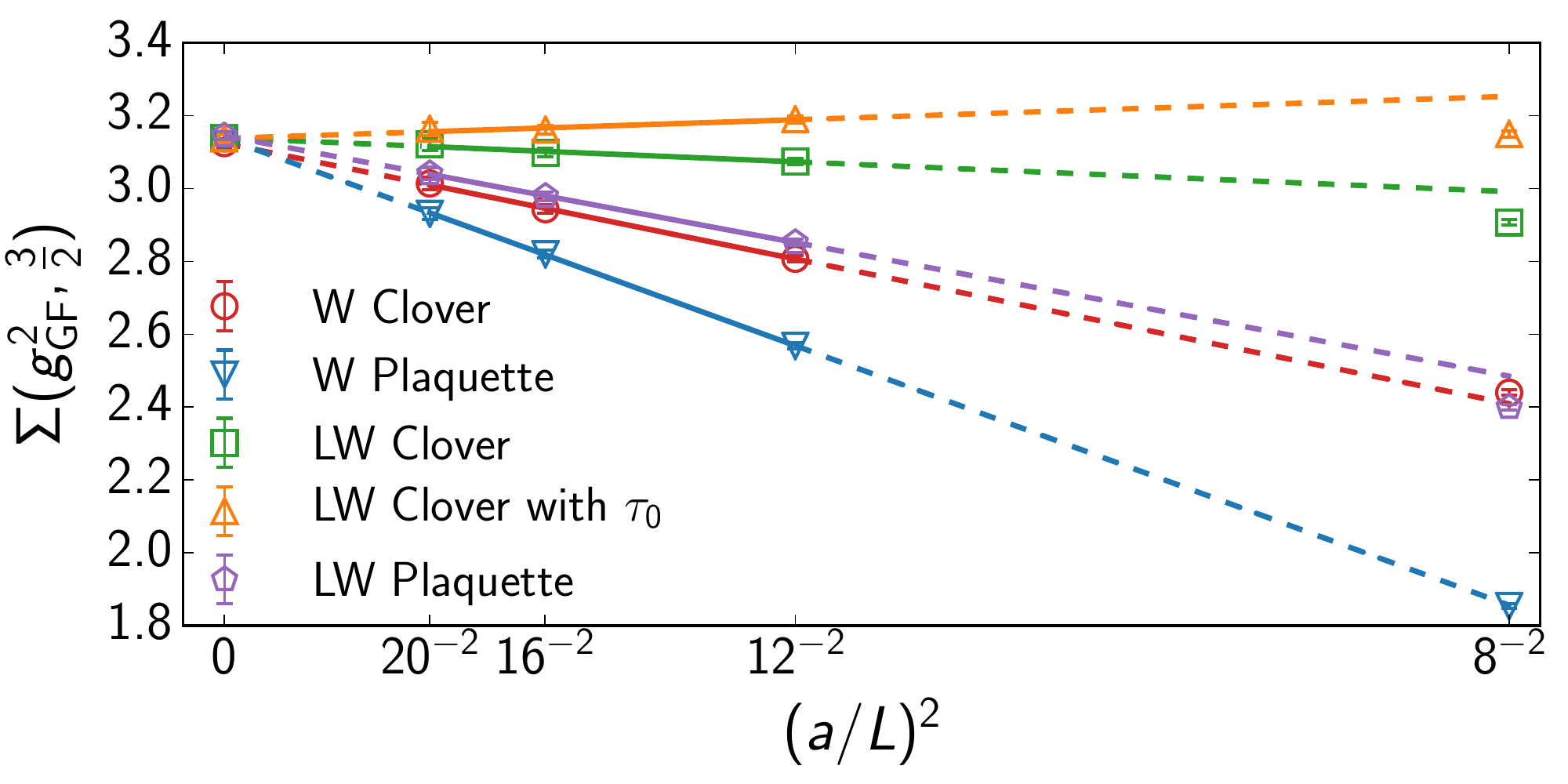}
  \includegraphics[width=8.6cm]{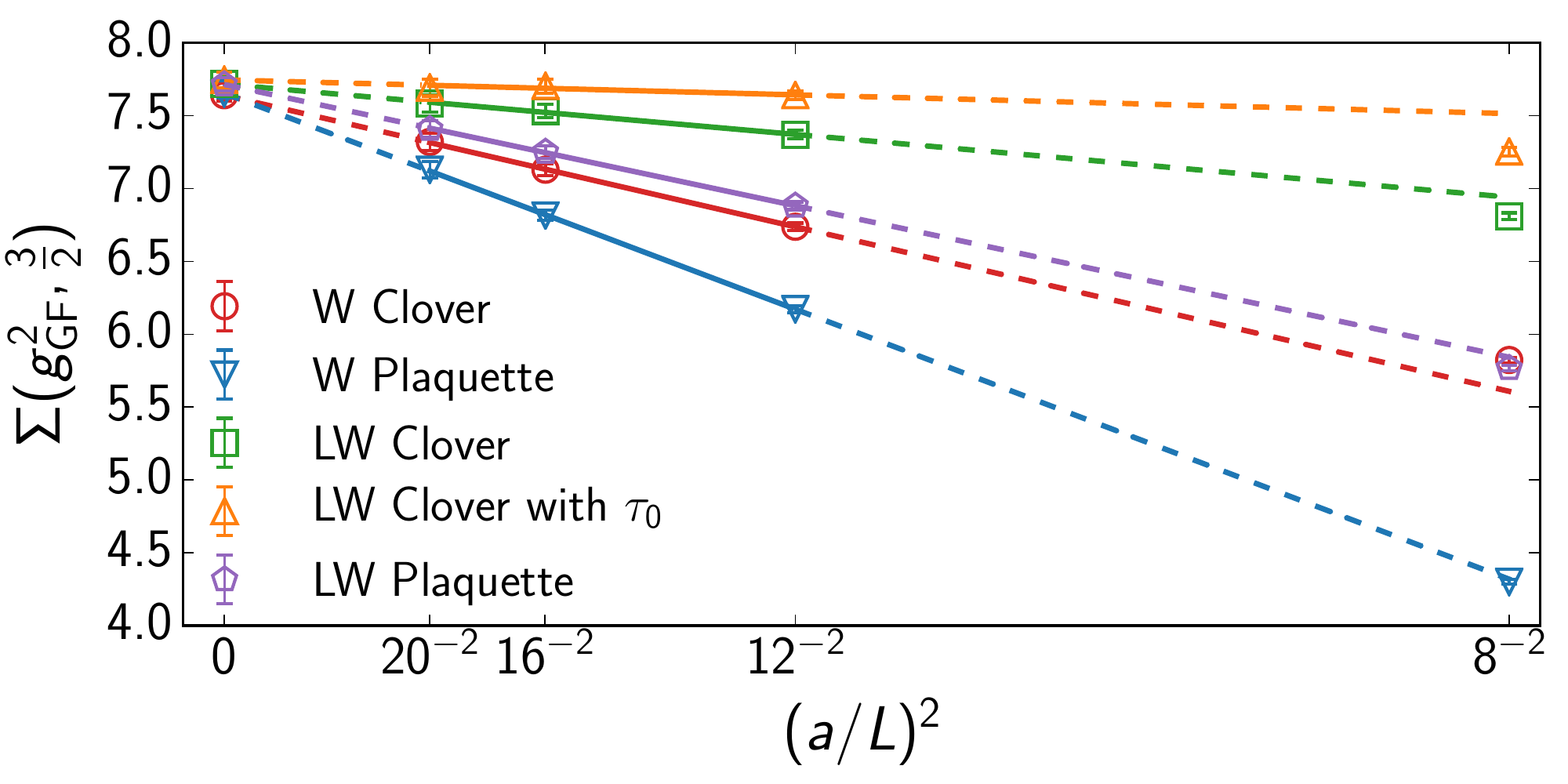}
  \includegraphics[width=8.6cm]{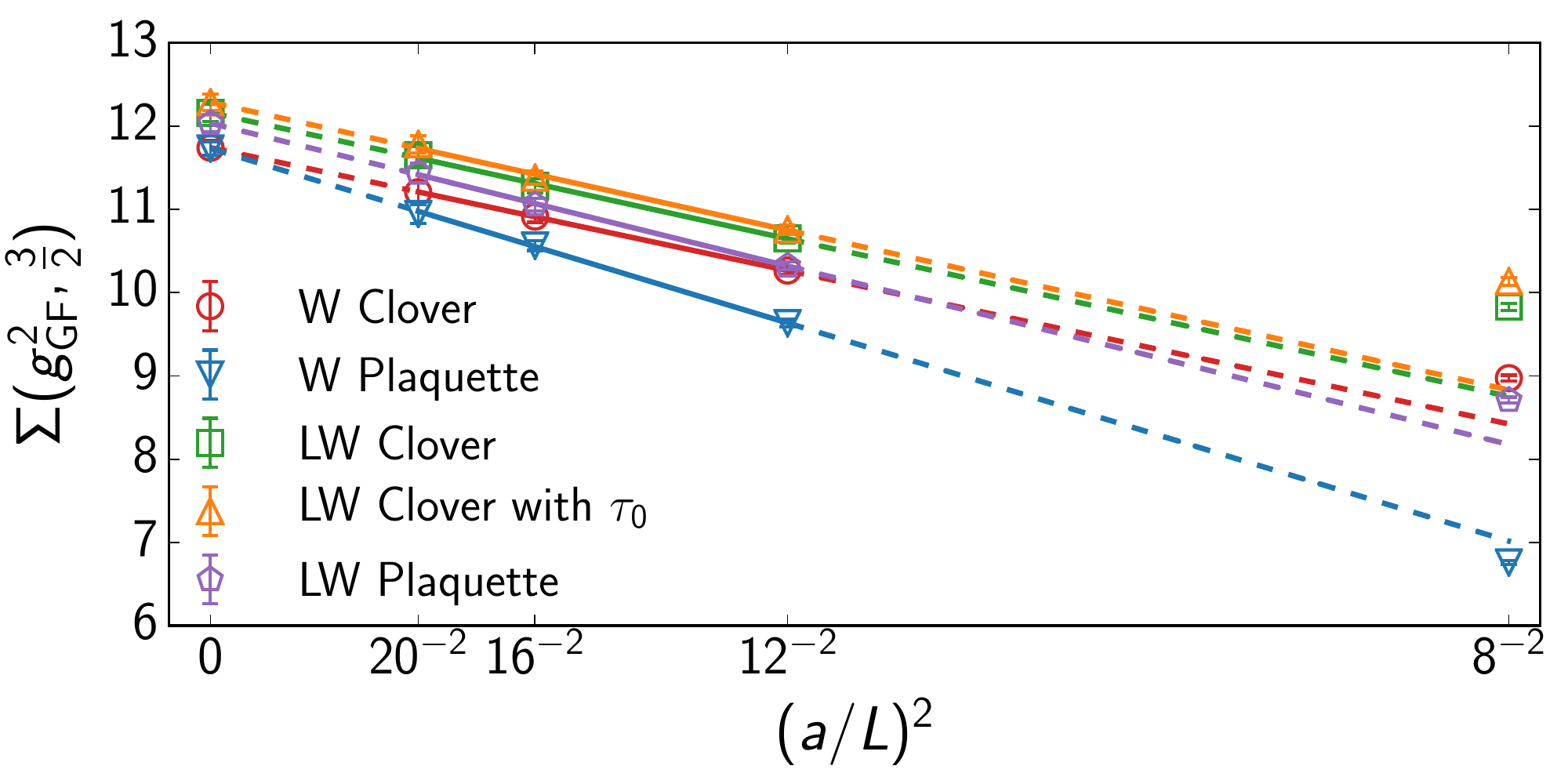}
  \includegraphics[width=8.6cm]{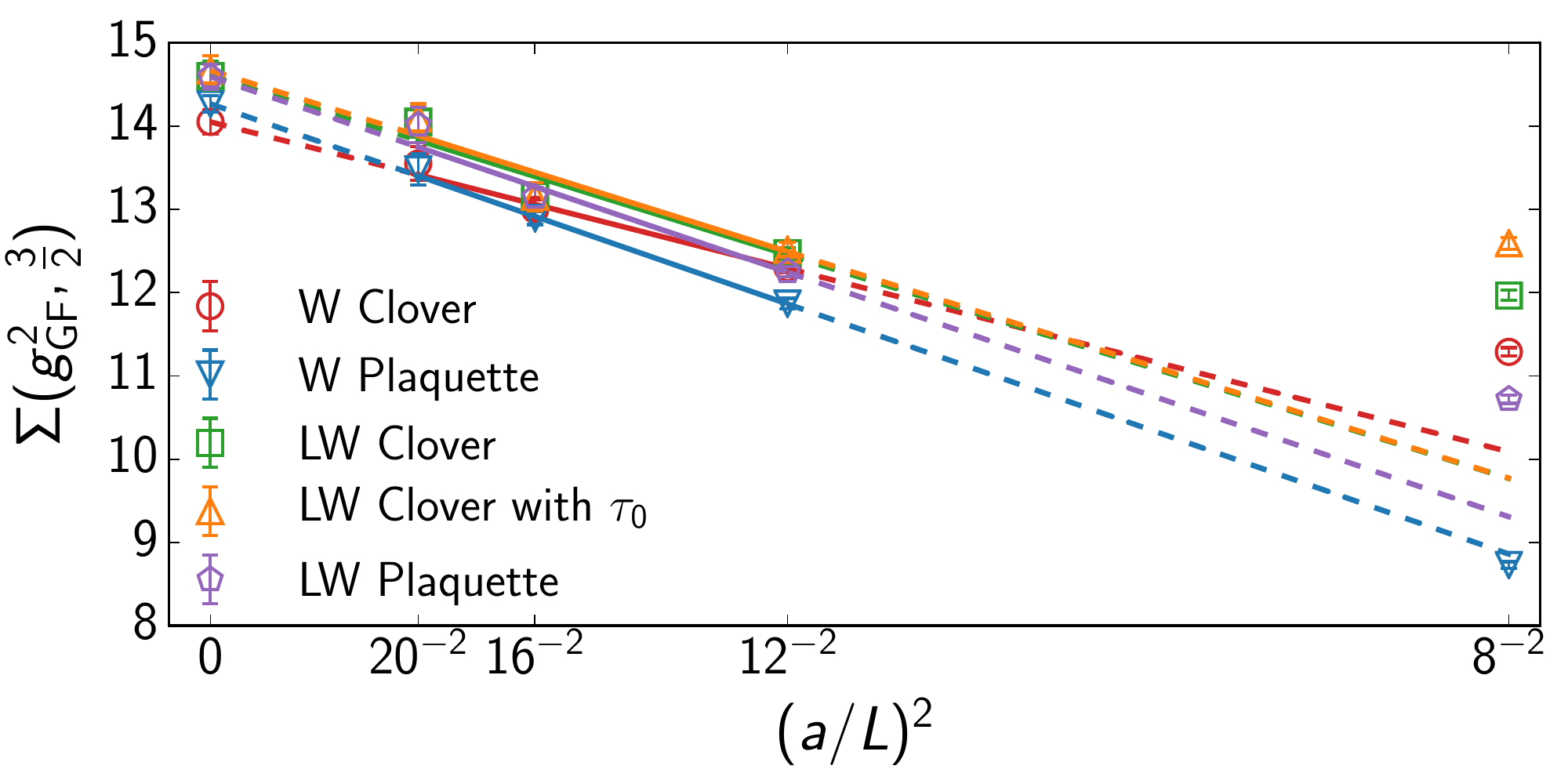}
  \caption[b]{ The effects of the choice of the Wilson (W) and the Lüscher-Weisz (LW) flow
    actions, the Clover and the Plaquette field strength observables and the
    $\tau_0$ correction~\eq{eq:taufunc} on the continuum limit of the
    step scaling function~\eq{eq:lat_step_cont} for $c_t=0.3$,
    measured at couplings (from top to bottom) $\gGF=3$, $\gGF=7$,
    $\gGF=11$ and $\gGF=14.5$.  }
\label{fig:contsigma}
\end{figure}%

In Fig.~\ref{fig:contsigma} we show the continuum limit extrapolation
of the step scaling function when $\gGF$ is varied from weak to strong
coupling, obtained using Luscher-Weisz or Wilson flow actions and
clover or plaquette field strength observables.  At small couplings
the continuum limit is very well under control: different
discretizations extrapolate very close to the same value.  At
couplings $\gGF \gsim 10$ the continuum limits start to show a few per
cent scatter.  This is taken into account in
the systematic uncertainties of the final results.

The $\tau_0$-correction parameter in Eq.~\eqref{eq:g2gf} can be tuned
to reduce most of the $O(a^2)$ errors from the continuum limit extrapolation of the step scaling 
function~\eq{eq:lat_step_cont}.  The parameter $\tau_0$ should have a small effect 
in the continuum extrapolation, so long as it is not too 
large~\cite{Hasenfratz:2014rna}.  In practice, we have observed that $\tau_0$ which depends logarithmically on $\gGF$ works well at small coupling \cite{Leino:2017lpc}.   With $c_t=0.3$, Luscher-Weisz flow action and clover field strength observable we use
\begin{equation}
  \tau_0 = 0.025\log(1+2\gGF)\,,
\label{eq:taufunc}
\end{equation}%
which makes the interpolation errors almost vanish at 
$\gGF \lsim 10$, as can be observed in Fig.~\ref{fig:contsigma}.  At larger couplings
$\tau_0$ correction cannot remove $O(a^2)$ significantly without ruining the continuum limit.
We note that in order to have consistent $\mathcal{O}(\A^2)$ shift in the step scaling analysis,
the $\tau_0$ correction should be a function of $\gGF$ instead of the bare coupling $g_0^2$~\cite{Ramos:2015dla}.
Because adjusting $\tau_0$ changes the value of the measured $\gGF$, the 
final value of $\tau_0$ is found by iterating equations \eqref{eq:g2gf}  and \eqref{eq:taufunc},  starting from the initial value $\gGF=g_0^2$.

\begin{figure}[t]
  \includegraphics[width=8.6cm]{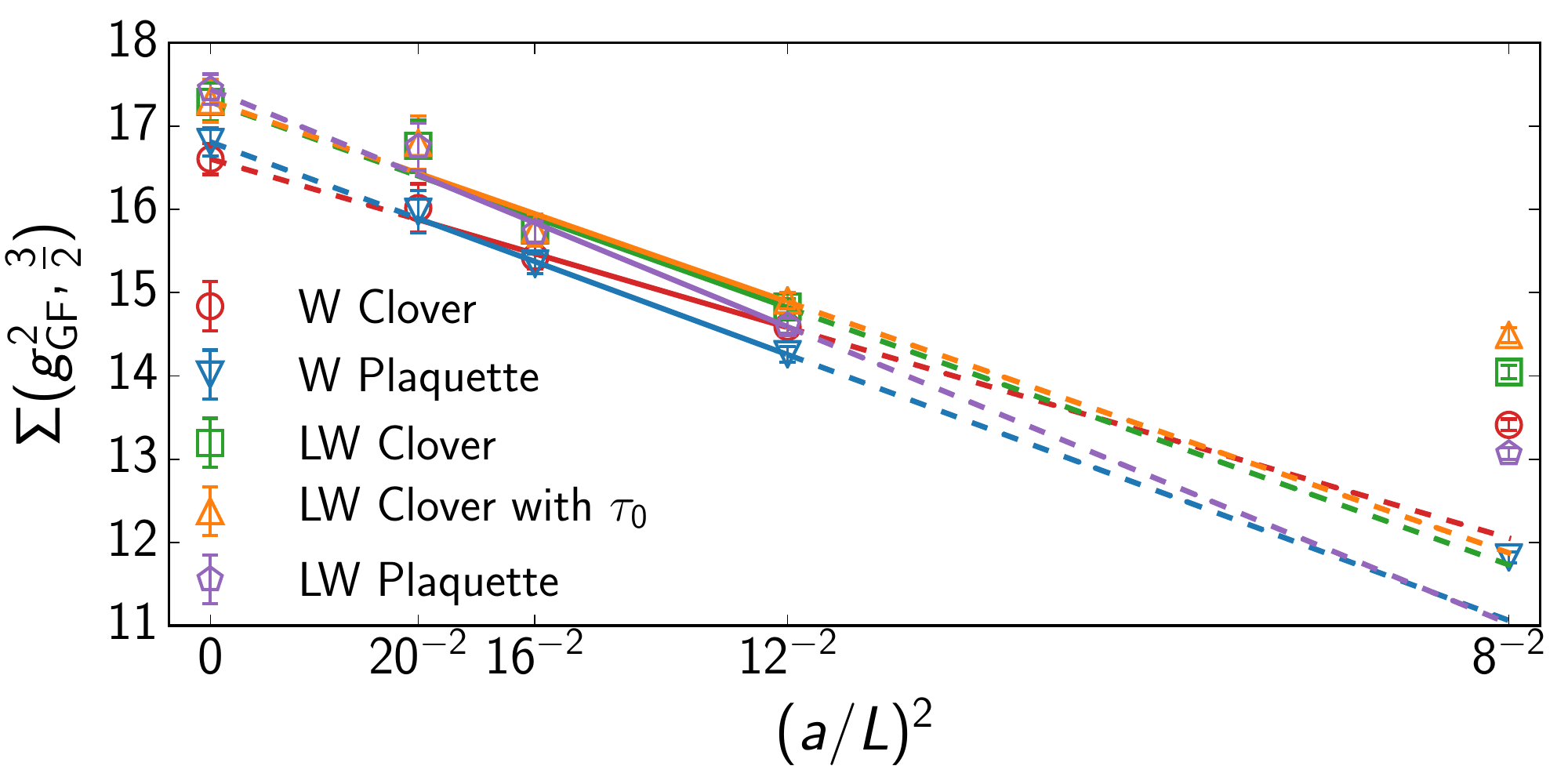}
  \includegraphics[width=8.6cm]{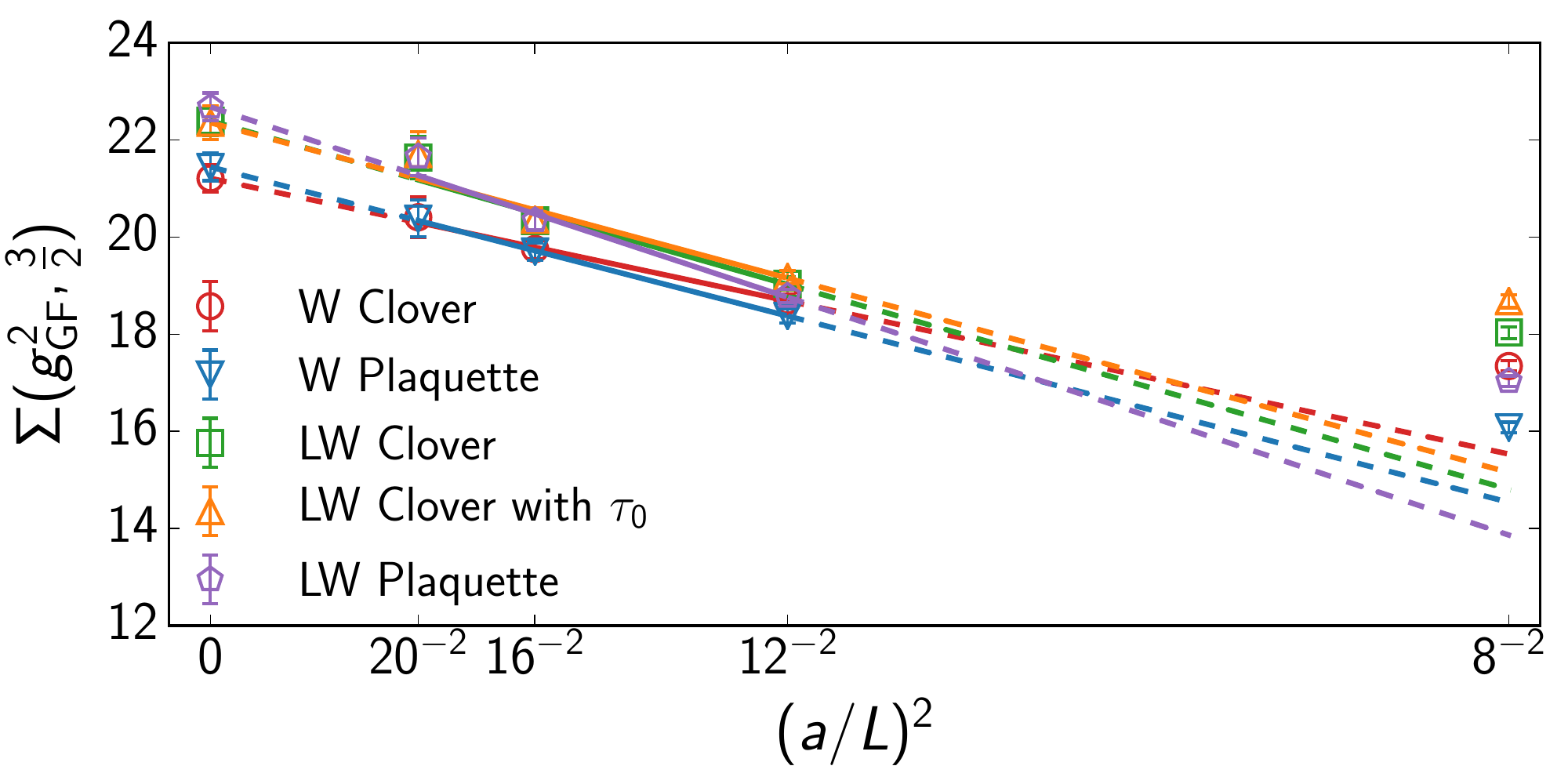}
  \includegraphics[width=8.6cm]{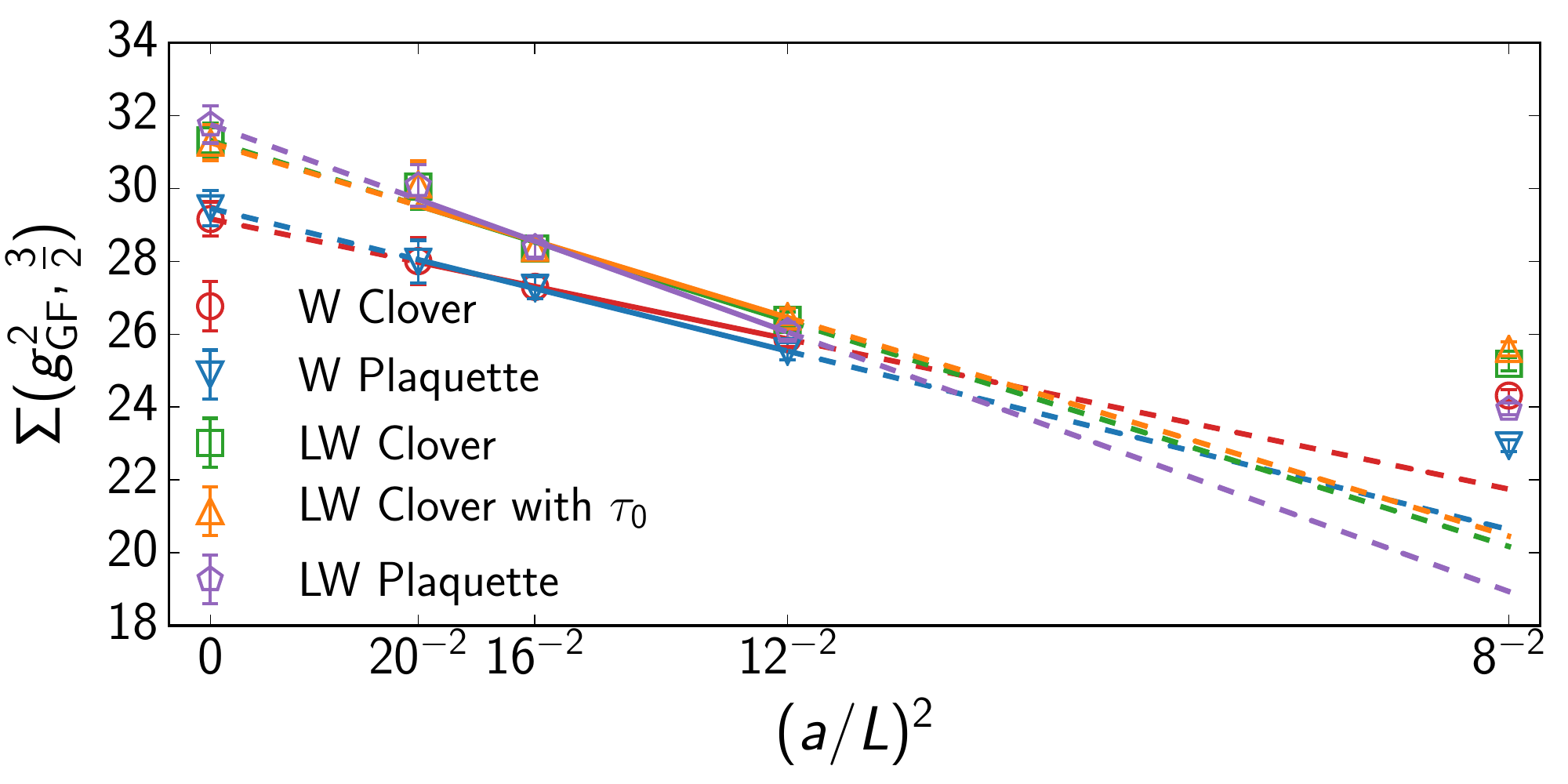}
  \caption[b]{
    {\em From top to bottom:} Same as Fig~\ref{fig:contsigma} but with: $c_t=0.35$,
    $c_t=0.4$ and $c_t=0.45$ measured at their respective IRFP's given in Table~\ref{tab:meas}
  }
\label{fig:contsigma_ct}
\end{figure}

In Fig.~\ref{fig:contsigma_ct} we show the continuum limit of the step scaling at $c_t = 0.35$, $0.4$ and $0.45$, evaluated at the IRFP of each $c_t$.  The couplings $\gGF = g_\ast^2$ at the fixed point are shown in Table \ref{tab:meas}. Because different $c_t$-values correspond to different coupling constant scheme, the values of $g_\ast^2$ vary significantly.  It is evident that as $c_t$ is increased the difference between the Lüscher-Weisz and Wilson flow actions grows, contributing to increasing systematic errors.

\begin{figure}[t]
  \includegraphics[width=8.6cm]{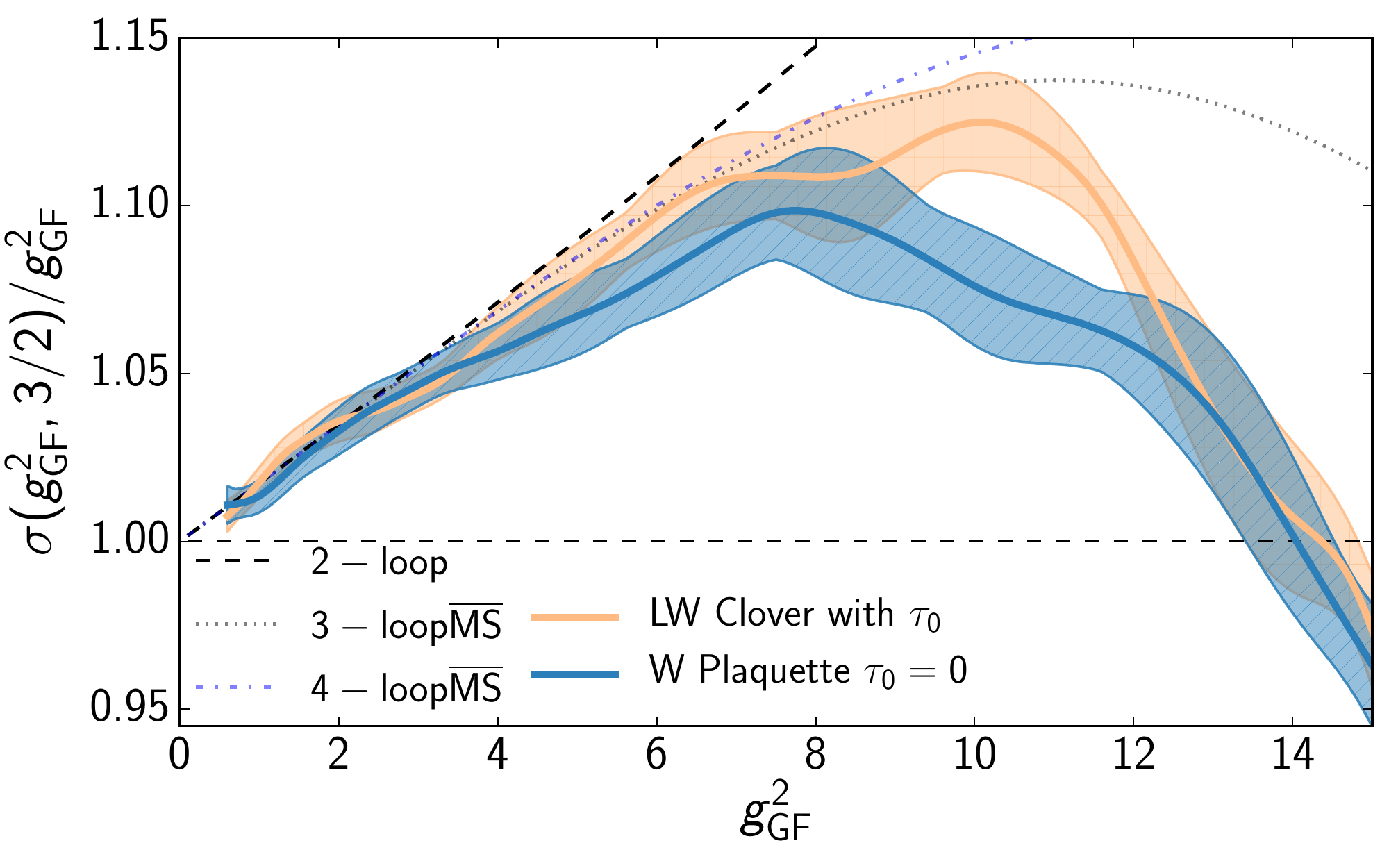}
  \includegraphics[width=8.6cm]{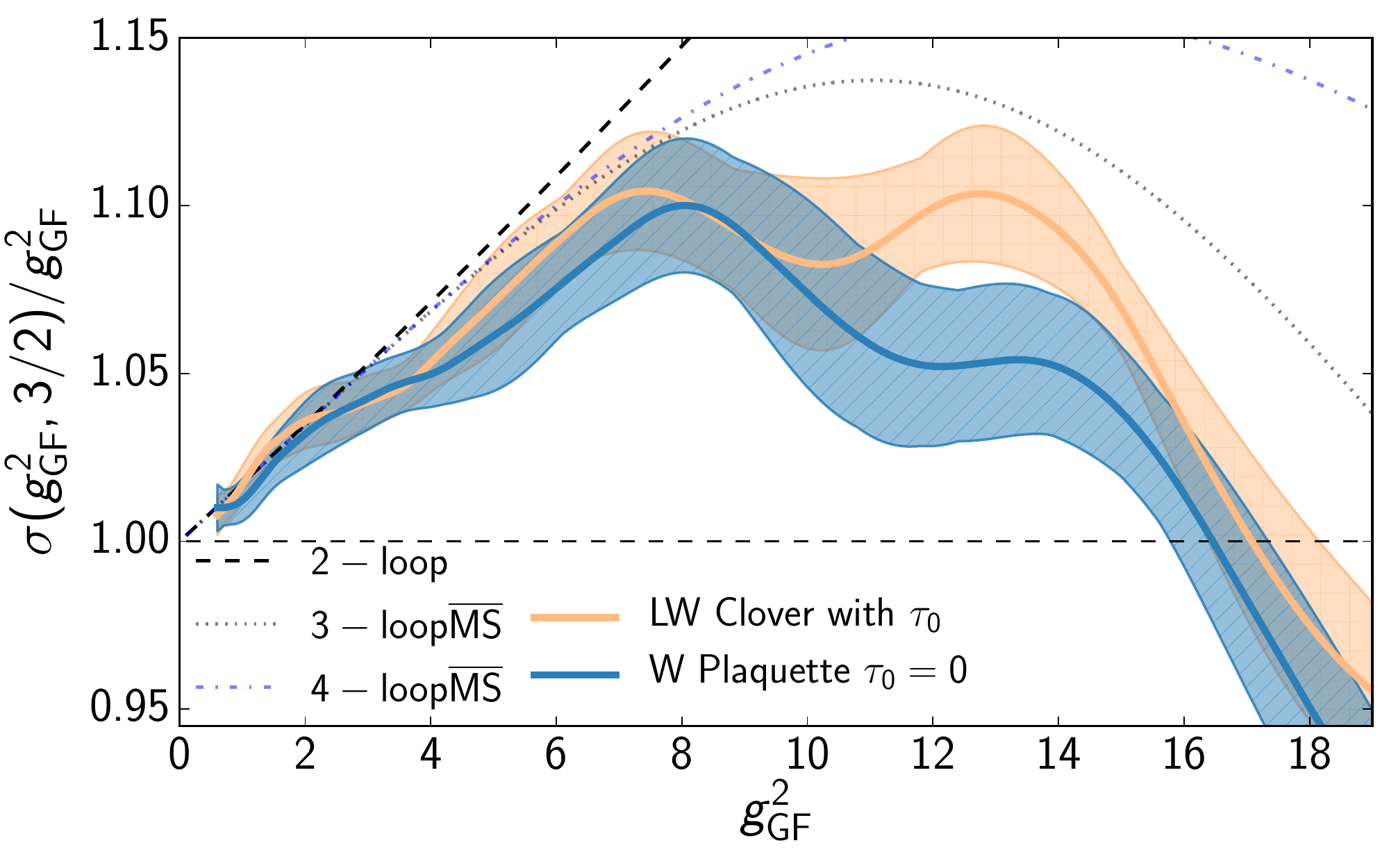}
  \includegraphics[width=8.6cm]{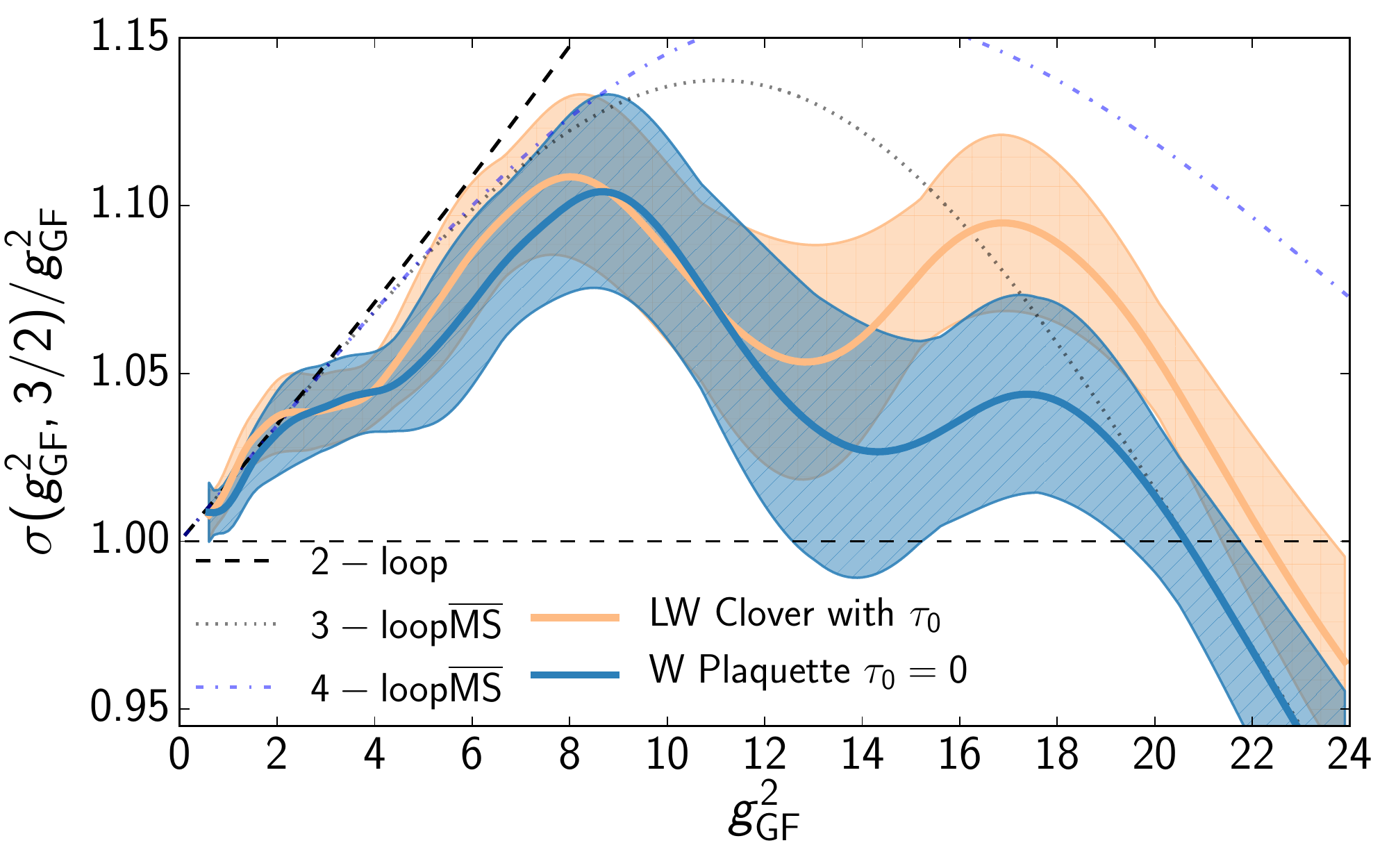}
  \caption[b]{
  			   {\em From top to bottom:} The continuum limit of the interpolated step scaling function
                           at $c_t=0.3$, $c_t=0.35$ and $c_t=0.4$, using
                           discretizations which give the smallest and largest result for the step scaling
                           function.
                         }
\label{fig:contsigma_full}
\end{figure}%

Finally, in Fig~\ref{fig:contsigma_full} we show the continuum limit of the step scaling across the full range of $\gGF$ at $c_t = 0.3$, $0.35$ and $0.4$.
The scheme independent 2-loop result and the scheme dependent 3- and 4-loop $\MSb$ results are shown as references,
while the 5-loop curve from Fig.~\ref{fig:g2_lat_step} is not shown here as it would mostly be outside the figure.
The error bands include the statistical errors and systematic uncertainty arising from different interpolating polynomials,~\eq{eq:betafitfun}.  
It is evident that as $c_t$ increases the reliability of the continuum limit extrapolation decreases, and
already at $c_t=0.4$ the final result has clearly unphysical strong ``wavy'' structure.  This is caused by the
use of the polynomial interpolation functions in Eq~\eqref{eq:betafitfun}.  We note that if we would
use polynomials of smaller degree (for example, $m=8$ in \eqref{eq:betafitfun}) the wavy structure would be strongly
reduced and error bands would be much narrower;  however, the $\chi^2/$d.o.f-values would not be acceptable.

Each value of $c_t$ corresponds to different coupling constant scheme,
and the value of the fixed point coupling strongly depends on the value of $c_t$.
The location of fixed point for each $c_t$ is reported in Table~\ref{tab:meas}
and individually for all discretizations of the flow in Table~\ref{tab:irfptable} in the appendix.
The first error is the statistical uncertainty
and the second error estimates systematic effects by including the full range of different discretization choices that were present in the Fig.~\ref{fig:contsigma}.
For our benchmark value $c_t=0.3$
we find that our continuum extrapolated results are compatible within $1\sigma$
level with respect to all these effects
except in the interval $\gGF \in [8,12]$ where LW and W evolved flows disagree slightly.
In this case the fixed point coupling has the value $g_\ast^2 = 14.5(4)_{-1.2}^{+0.4}$

\begin{table}
\begin{ruledtabular}
\begin{tabular}{ccccc}
 & $c_t=0.3$ & 0.35 & 0.4 & 0.45 \\
\hline
$g_\ast^2$      & $14.5(4)_{-1.2}^{+0.4}$  & $17.1(5)_{-1.3}^{+0.8}$  & $22.2(6)_{-2.5}^{+1.3}$  & $31(1)_{-18}^{+2}$     \\
$\gamma_g^\ast$ & $0.648(97)_{-0.1}^{+0.16}$ & $0.71(12)_{-0.11}^{+0.17}$ & $0.73(10)_{-0.18}^{+0.11}$ & $0.75(12)_{-0.61}^{+0.12}$ \\
\end{tabular}
\end{ruledtabular}
\caption{\label{tab:meas}
Measured couplings and critical exponents with different choices of parameter $c_t$.
The error shown in parenthesis is the statistical uncertainty, and the super- and subscripts are the systematic errors due to different discretizations of the gradient flow and the field strength observables.}

\end{table}%

\section{Leading irrelevant critical exponent}\label{sec:gammag}
We can also obtain the leading irrelevant exponent $\gamma_g^\ast$ at the fixed point, defined by the slope of the $\beta$-function at the IRFP.  This quantity is scheme independent, and thus should not depend on $c_t$.

In the proximity of the fixed point we can approximate the $\beta$-function as
\begin{align}
\beta(g) &= -\mu \frac{dg^2}{d\mu} \approx \gamma_g^\ast(g^2-g_\ast^2) \label{eq:beta*}\\
&\approx \bar\beta(g) \equiv
\frac{g}{2\ln(s)} \left ( 1 - \frac{\sigma(g^2,s)}{g^2} \right ).\nonumber
\end{align}%
Measuring the slope of the step scaling function $\sigma(g^2)$ around the fixed point gives the exponent 
$\gamma_g^\ast=0.648(97)_{-0.1}^{+0.16}$ 
at $c_t=0.3$; the results with other $c_t$ are shown in Table \ref{tab:meas}.
While there is noticeable variance between different discretizations, as indicated by the second set of errors,
the result is compatible with the recent scheme independent estimate of $\gamma_g^\ast=0.6515$
in Refs.~\cite{Ryttov:2017kmx,Ryttov:2017toz}.
The results obtained with different discretizations are shown individually in Table~\ref{tab:irfpstar}.
When $c_t$ is varied, the value of $\gamma_g^\ast$ remains constant within errors, 
in accord with the scheme independence of this quantity.

The results obtained above rely on the accurate continuum limit of the step scaling function $\sigma(g^2)$. 
However, as discussed in section \ref{sec:evolution}, 
the continuum limit may be in effect somewhat different at the close proximity of the IRFP.  
We can verify the consistency of the results by using a finite size scaling method 
developed in~\cite{Appelquist:2009ty,DeGrand:2009mt,Lin:2015zpa,Hasenfratz:2016dou} 
to get an alternative measurement of $\gamma_g^\ast$.  
In the close proximity of the IRFP, by integrating~\eqref{eq:beta*} 
we obtain a finite size scaling relation between lattices of size $L_\mathrm{ref}$ and $L$~\cite{Lin:2015zpa}:
\begin{equation}
\gGF(\beta_L,L) - g_\ast^2 = 
\left[ \gGF(\beta_L,L_\mathrm{ref})-g_\ast^2\right]\left(\frac{L_\mathrm{ref}}{L}\right)^{\gamma_g^\ast}
\label{eq:ramosgamma}
\end{equation}
This equation relies on the evolution of the coupling towards the fixed point as the lattice size
is increased from $L_\mathrm{ref}$ to $L$.  Hence, it cannot be used exactly at the fixed point where
there is no evolution, but only in some environment around it.
We note that this also assumes vanishing discretization artifacts, 
and thus it can be used only if the lattices are already close enough to the continuum ($L$ large).  

\begin{figure}[t]
  \includegraphics[width=8.6cm]{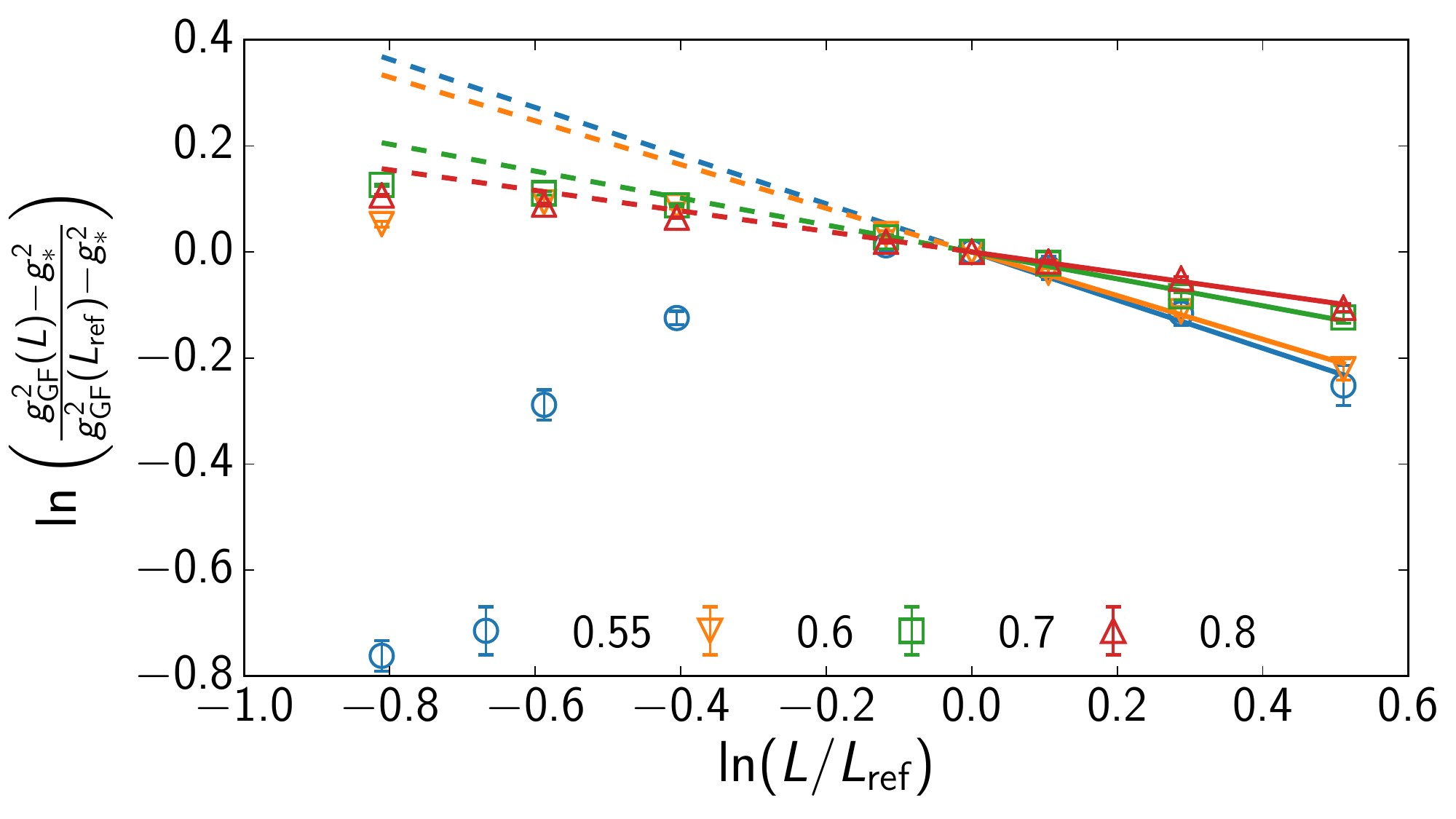}
  \caption[b]{
    Fit to function~\eqref{eq:ramosgamma} for measured couplings $\gGF(\beta_L,L)$ at $\beta_L=  0.55 \ldots 0.8$ at $c_t=0.3$, using
                           $L_\mathrm{ref}/a = 18$.
  }
\label{fig:gammaglines}
\end{figure}

In Fig.~\ref{fig:gammaglines} we show the fit to Eq.~\eqref{eq:ramosgamma} to individual measurements
of $\gGF$ at $\beta_L \le 0.8$, corresponding to measurements which are close to the fixed point.
A good fit to Eq.~\eqref{eq:ramosgamma} is obtained if we choose $L_\mathrm{ref}/a \ge 18$,
allowing us to extract an estimate for $\gamma_g^\ast$.

\begin{figure}[t]
  \includegraphics[width=8.6cm]{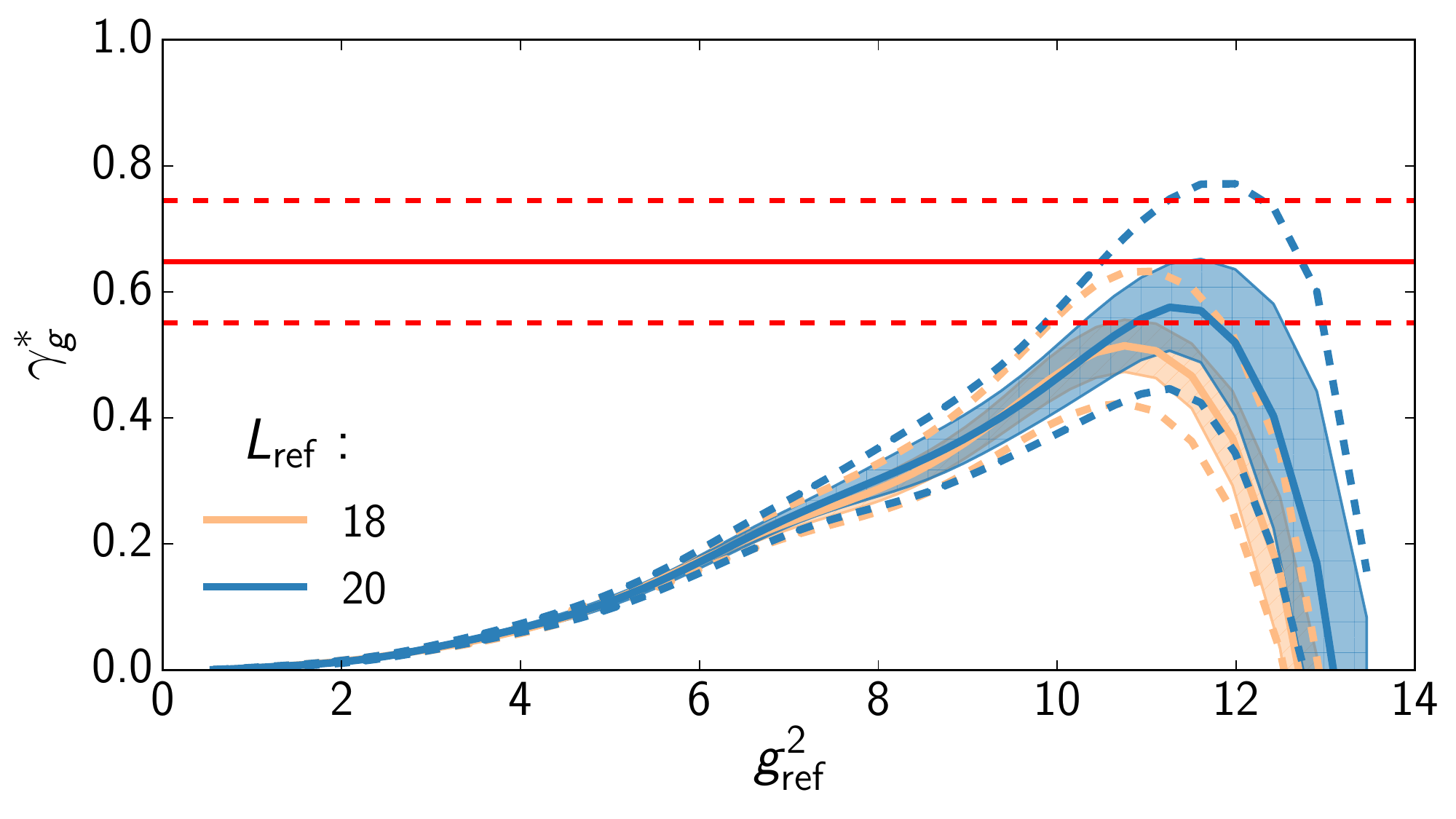}
  \includegraphics[width=8.6cm]{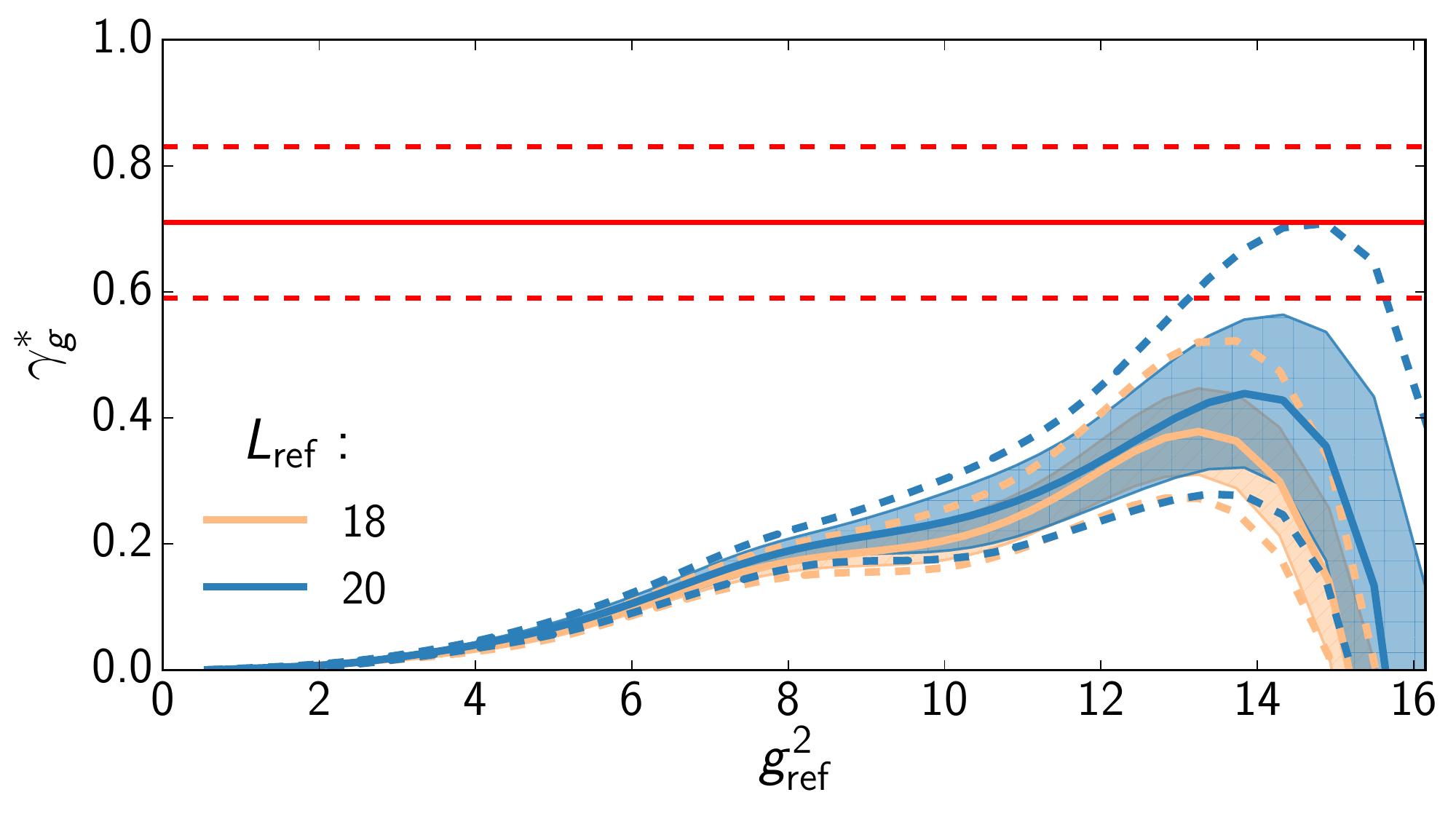} 
  \caption[b]{
  			   Fit to function~\eqref{eq:ramosgamma} for all couplings $\gGF(\beta_L,L_\mathrm{ref})$
			   with chosen set of discretizations:
  			   {\em Top:} at $c_t=0.3$, and {\em Bottom:} at $c_t=0.35$.
			   The shaded bands indicate the statistical errors for $g_\ast^2$ being the measured value,
			   and dashed lines indicate how the result changes when $g_\ast^2$ is varied within
			   its statistical errors.
  }
\label{fig:ramos1}
\end{figure}

Instead of using individual measurements we use the interpolated values of $\gGF(\beta_L,L)$, because
this allows us to freely tune the value of $\beta_L$.  
In Fig~\ref{fig:ramos1} we show the resulting $\gamma_g^\ast$ from fits to Eq.~\eqref{eq:ramosgamma},
plotted as functions of $g^2_\mathrm{ref}\equiv \gGF(\beta_L,L_\mathrm{ref})$.
The red lines correspond to the values given in Table~\ref{tab:irfpstar}, measured from the slope of the step scaling function.
The shaded error bands correspond
to statistical errors while keeping the values of $g_\ast^2$ fixed to central values in Table~\ref{tab:irfptable},
and the dashed lines show the variation of the result if we allow $g_\ast^2$ to vary within the statistical error range.

The resulting $\gamma_g^\ast$ is expected to be close to the true $\gamma_g^\ast$ only in close proximity of the
IRFP.  However, too close to the IRFP Eq.~\eqref{eq:ramosgamma} becomes unstable,
which is indicated by a sudden drop in the $\gamma_g^\ast$ measurements.  Indeed, at $g_\mathrm{ref}^2 \approx 12$ we
observe the $c_t=0.3$ case to give $\gamma_g^\ast$ which is in agreement with the one obtained from the slope
of the $\beta$-function.   At small $g^2_\mathrm{ref}$ the measurement of $\gamma_g^\ast$ using Eq.~\ref{eq:ramosgamma}
approaches zero.

\vspace{2mm}
\section{Anomalous dimension of the mass operator}
\label{sec:anomalous}
\begin{figure}
\begin{center}
\includegraphics[width=8.6cm]{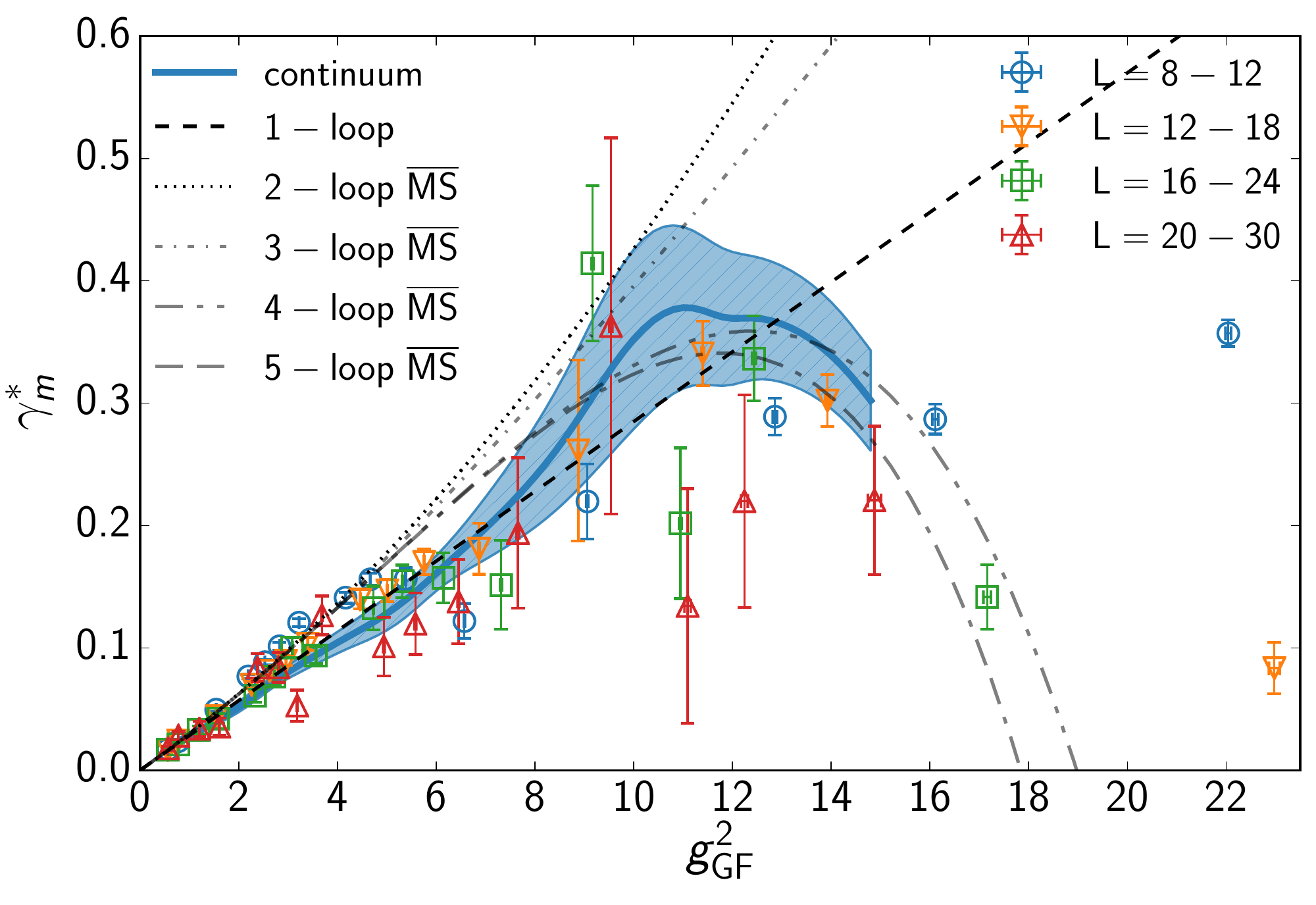}
\end{center}
\caption{The mass anomalous dimension as a function of the gradient flow coupling constant obtained using the mass step scaling function
		 and its continuum limit.
		 The results become unstable at large couplings.}
\label{fig:gamma1}
\end{figure}

In order to measure the anomalous dimension of the fermion mass operator $\gamma_m^\ast$ we use two different methods,
the mass step scaling method and the spectral density method.
In the step scaling method we measure $\gamma_m$
from the running of the pseudoscalar density renormalization constant~\cite{Capitani:1998mq,DellaMorte:2005kg}
\begin{align}
Z_P(g_0,L) = \frac{\sqrt{2 f_1} }{f_P(L/2)}\,,
\label{Zp}
\end{align}%
where $f_P$ and $f_1$ are pseudoscalar current densities defined explicitly in e.g.~\cite{Luscher:1996vw,Leino:2017lpc}.
The mass step scaling function is defined as~\cite{Capitani:1998mq}:
\begin{align}
\Sigma_P(u,s,L/a) &=
   \left. \frac {Z_P(g_0,sL/a)}{Z_P(g_0,L/a)} \right |_{g_{\rm{GF}}^2(g_0,L/a)=u}
   \label{Sigmap}
\end{align}%
As in the case of the coupling, we choose $s = 3/2$.
The continuum limit $\sigma_P(u,s)$ is obtained by interpolating the measured $Z_p$ by
8th order polynomials and assuming $\mathcal{O}(a^2)$ errors.
The mass anomalous dimension is then obtained as~\cite{DellaMorte:2005kg}
\begin{align}
  \gamma_m^\ast(u) = -\frac{\log \sigma_P(u,s)}{\log s }\,.
\label{eq:gammastar}
\end{align}%
The results are shown in Fig.~\ref{fig:gamma1} and the raw data is given in Table~\ref{tab:zp_sup11}.
The method gives results comparable to one loop perturbation theory predictions at small gauge coupling $\gGF$.
While the higher loop $\MSb$ expansions~\cite{Vermaseren:1997fq,Luthe:2016xec} 
are scheme dependent and cannot be directly compared to our results,
it is nevertheless comforting to observe comparable behaviour between our result to the 4 and 5-loop behavior.
However, the method becomes unstable at large coupling, which implies that
at the fixed point $g_\ast^2 \approx 14.5$ the continuum limit cannot be trusted.

The second way to measure $\gamma_m$ is based on the fact that the it also determines the
scaling of the spectral density of the massless Dirac operator.
The explicit calculation of the eigenvalue distribution is prohibitively costly,
but the stochastic methods~\cite{Giusti:2008vb}
have made it possible to determine the mass anomalous dimension from the
scaling of the mode number of the Dirac operator~\cite{Patella:2011jr}.
The mode number is known to follow a scaling behavior
\begin{equation}\label{modenumber1}
\nu(\Lambda) \propto \Lambda^{4/(1+\gamma_m^\ast)}\,,
\end{equation}%
in some energy range between the infrared and the ultraviolet in the vicinity of a fixed point.
Here $\gamma_m^\ast$ is the mass anomalous dimension $\gamma_m$ at the fixed point.

We calculate the mode number per unit volume of~\eq{modenumber1} by using
\begin{equation}
\nu(\Lambda) =\lim_{V\to \infty}  \frac{1}{V}\langle \textnormal{tr }\mathbb{P}(\Lambda)\rangle\,,
\end{equation}%
where the operator $\mathbb{P}(\Lambda)$ projects
from the full eigenspace of $M = m^2 - \slashed{D}^2$ to the eigenspace of eigenvalues smaller than $\Lambda^2$.
The trace is evaluated stochastically~\cite{Giusti:2008vb},
and fitted to the power law behavior of \eq{modenumber1}.
However, the energy range where this power law behavior holds is not known beforehand,
and needs to be determined by observing the quality of the fit in a given range.

We use $L/a=24$ lattices from the step scaling analysis,
and take 12 to 20 well separated configurations for each value of the gauge coupling.
We calculate the mode number for 90 values of $\Lambda^2$ ranging from $10^{-3}$ to $0.3$.
The results are then fitted to~\eq{modenumber1}. 
The fit range is determined by varying its lower 
and the upper limits and observing the stability and the quality of the fit.
As a cross reference at weak coupling, the fitted value of
$\gamma_m^\ast$ and the value obtained with the step scaling method are compared.

\begin{figure}[t]
\begin{center}
\includegraphics[width=8.6cm]{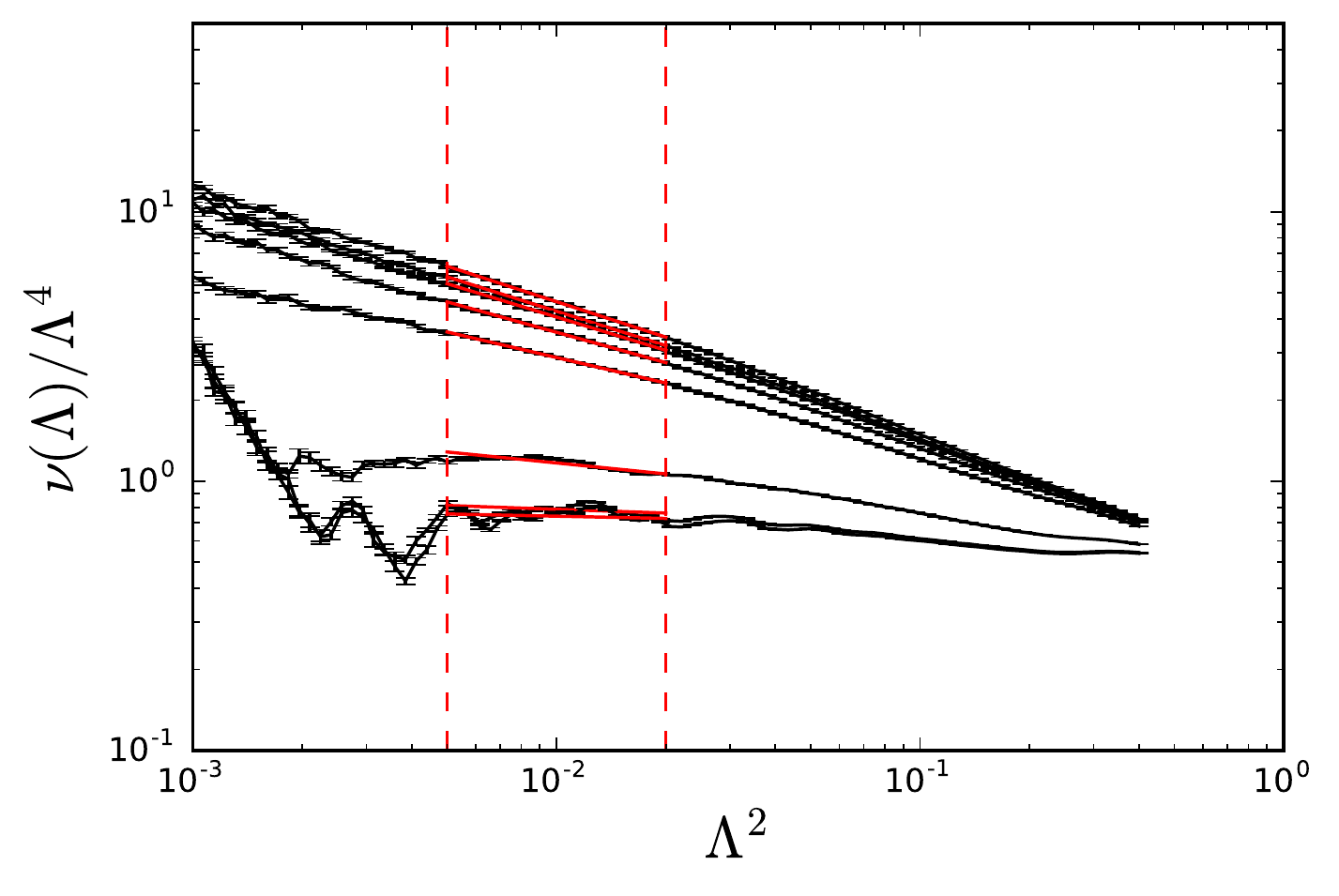}
\caption{The mode number divided by $a^4\Lambda^4$ as a function of $a^2\Lambda^2$ on a $L/a=24$ lattice.
         The dashed red lines indicate the chosen fit range and the red solid lines the fit function.
		 The fit ranges were varied around these chosen regions. The curves are in a descending gauge coupling order.}
\label{fit_range}
\end{center}
\end{figure}

In Fig.~\ref{fit_range} we plot the mode number divided by the fourth power of the eigenvalue scale,
where the the fit range and the fit function of~\eq{modenumber1} are shown overlaid in red.
According to~\eq{modenumber1} in the proximity of the fixed point the infrared behavior
should be a power law in the absence of lattice artifacts.
We observe this at strongest couplings,
however, at small couplings the low eigenvalues appear in discrete energies,
which manifests as the bumps in the mode number curve, making the power law less evident.
To illustrate the evolution of the mass anomalous dimension we use the same fit range for both weak and strong couplings.

\begin{figure}[t]
\begin{center}
\includegraphics[width=8.6cm]{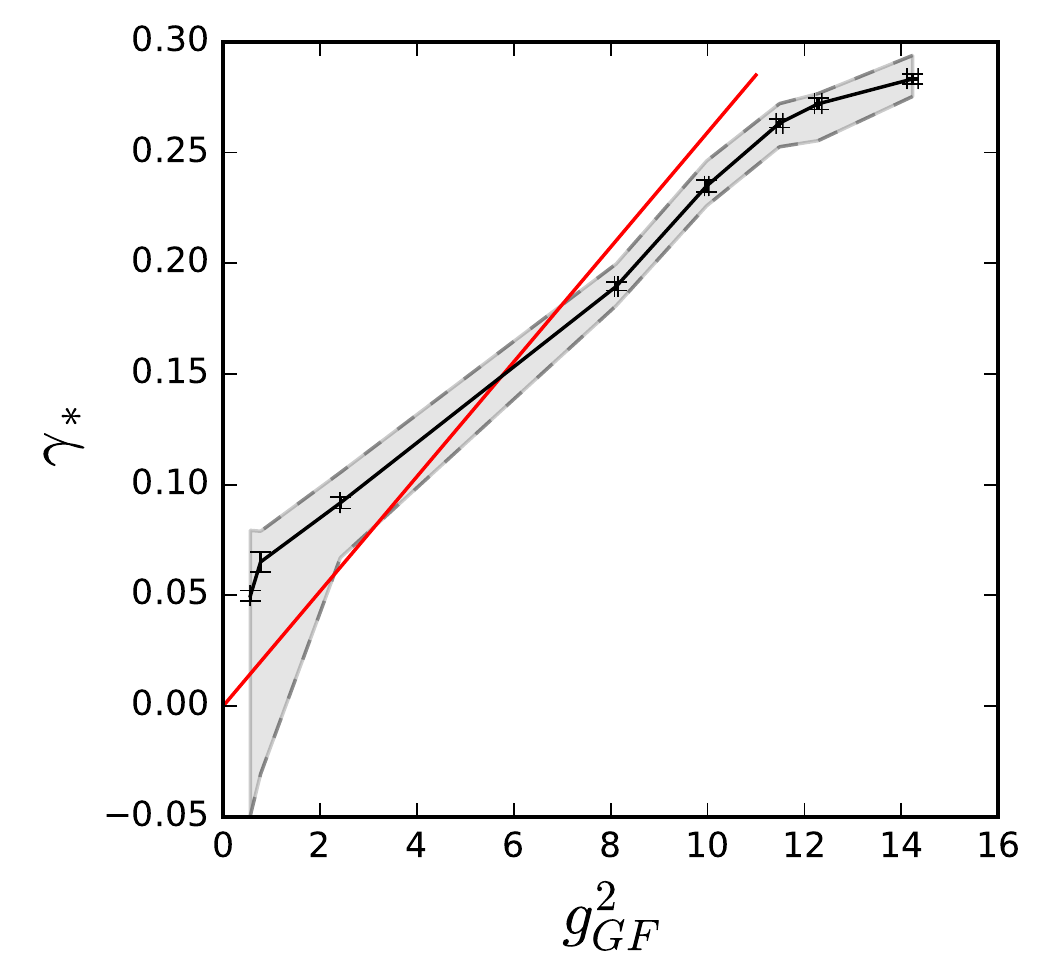}
\caption{The value of $\gamma_m^\ast$ obtained by fitting Eq.~\eqref{modenumber1} to the data in Fig.~\ref{fit_range}
         is shown with black points and the one loop perturbative result with a red line.
		 The shaded regions are estimates for reasonable ranges of values obtainable using the method,
		 and were obtained by varying the fit range shown in Fig.~\ref{fit_range} slightly.}
\label{result}
\end{center}
\end{figure}

The final result of the spectral density method is shown in Fig~\ref{result},
where the mass anomalous dimension $\gamma_m^\ast$, 
obtained by fitting the data with~\eq{modenumber1},
is shown as a function of the gauge coupling $\gGF$.
The shaded band illustrates the uncertainty resulting from varying the upper and lower limits of the fit range by $\sim 50$\%.
The largest uncertainty arises at small gauge couplings,
where the bumps in the data cause the changes in the fit range to change the fit dramatically.
The error band of Fig.~\ref{result} becomes narrower towards the larger couplings
as the ensembles near the IRFP are less sensitive to variations of the fit range.

At the fixed point $g_\ast^2=14.5$ we obtain $\gamma_m^\ast = 0.283(2)^{+0.01}_{-0.01}$.
However, this result is obtained at fixed lattice size $L/a=24$.
A proper continuum limit requires extrapolation to infinite $L$,
but at smaller $L/a$ the finite size effects make the usable range for the power law fit too narrow.

Interestingly, the mass step scaling method
and the spectral density method complement each other:
while the mass step scaling is stable and accurate at weak couplings,
where the spectral density method fails, at strong coupling the roles are reversed.

\vspace{2mm}
\section{Conclusions}
\label{sec:conclude}
We have studied the running coupling in the SU(2) lattice gauge theory
with 6 fermions in the fundamental representation. Gradient flow algorithm with
Dirichlet boundaries was shown to provide robust results on the large coupling behavior of this
theory giving a result consistent with 
the existence of IRFP at $g_\ast^2 = 14.5(4)_{-1.2}^{+0.4}$ 
in our benchmark scheme. The scheme-independent slope of $\beta$-function at IRFP was measured to be
$\gamma_g^\ast=0.648(97)_{-0.1}^{+0.16}$. 

We also determined the mass anomalous dimension $\gamma_m$ in this theory
using the spectral density method and the mass step scaling method.
The step scaling method gives results compatible with perturbation
theory at weak coupling. At intermediate couplings the step scaling method
can be matched on the results from the spectral density method
which remain stable in the vicinity of the fixed point
where the step scaling computation breaks down.
With the spectral density method we estimated the mass anomalous dimension at
the fixed point as $\gamma_m^\ast = 0.283(2)^{+0.01}_{-0.01}$, albeit a proper continuum limit is still lacking.

Our results are consistent with the existence of a strong coupling IRFP and 
indicate that the SU(2) gauge theory with six fermion
flavors in the fundamental representation is within the conformal window.
The theory likely is near the lower boundary of the conformal window,
which makes it an interesting candidate for beyond the standard model theories:
when coupled with the electroweak gauge currents the chiral symmetries are explicitly broken,
the theory is pulled outside the conformal window and may constitute a concrete
example of a walking technicolor theory.

\FloatBarrier
\acknowledgments{
This work is supported by the Academy of Finland grants 267286 and 267842.
V.L and J.M.S. are supported by the Jenny and Antti Wihuri foundation
and S.T. by the Magnus Ehrnrooth foundation.
The simulations were performed at the Finnish IT Center for Science (CSC), Espoo, Finland.
}

\bibliography{su2_nf6.bib}{}
\bibliographystyle{apsrev4-1.bst}

\appendix
\onecolumngrid
\section{Tables}

\begin{table}[b]
\caption{\label{tab:kappa_sup1}
The measured $\kappa_c(\beta_L)$ at $L/a=24$ for each $\beta_L$.
}
\begin{ruledtabular}
\begin{tabular}{cccccccc}
$\beta_L$ & $\kappa_c $ & $\beta_L$ & $\kappa_c $ & $\beta_L$ & $\kappa_c $ & $\beta_L$ & $\kappa_c $ \\
\hline
8    & 0.125310366353981 & 2   & 0.127533813721664 & 1   & 0.131448889150607 & 0.6  & 0.136438136224601 \\
6    & 0.125459579958083 & 1.7 & 0.128194200995596 & 0.9 & 0.132331360707040 & 0.55 & 0.137424583321490 \\
4    & 0.125860459184944 & 1.5 & 0.128799165934744 & 0.8 & 0.133419041876613 & 0.53 & 0.137839481272905 \\
3    & 0.126367585261215 & 1.3 & 0.129603737388233 & 0.7 & 0.134765027707880 & 0.5  & 0.138504981089103 \\
\end{tabular}
\end{ruledtabular}
\end{table}

\begin{table}[b]
\caption{\label{tab:trajnum_sup2}
Number of trajectories for each $\beta_L$ and $L$ after thermalization.
}
\begin{ruledtabular}
\begin{tabular}{ccccccccc}
$\beta_L$ & $N(L=8)$ & $N(L=10)$ & $N(L=12)$ & $N(L=16)$ & $N(L=18)$ & $N(L=20)$ & $N(L=24)$ & $N(L=30)$ \\
\hline
8    &  81351  & 10849 & 78537  & 8500   & 6468   & 11473 & 62574 & 7383  \\
6    &  157185 & 20468 & 89006  & 122197 & 95460  & 40434 & 33845 & 6098  \\
4    &  95516  & 20604 & 84883  & 106793 & 86888  & 41198 & 14031 & 5963  \\
3    &  101614 & 23139 & 88269  & 102191 & 82956  & 39127 & 21475 & 8520  \\
2    &  94905  & 17527 & 82783  & 94976  & 76712  & 35925 & 40449 & 9146  \\
1.7  &  93581  & 19990 & 79821  & 92194  & 74062  & 34220 & 36175 & 8785  \\
1.5  &  92038  & 19268 & 113427 & 90364  & 70024  & 32955 & 21173 & 10895 \\
1.3  &  89055  & 18380 & 110383 & 88057  & 69042  & 31553 & 32209 & 12014 \\
1    &  85016  & 16659 & 105548 & 75659  & 75037  & 33030 & 19082 & 11730 \\
0.9  &  100759 & 22780 & 106021 & 77452  & 72799  & 46582 & 47578 & 15254 \\
0.8  &  78037  & 29807 & 135876 & 95623  & 97127  & 71468 & 42482 & 21425 \\
0.7  &  130058 & 30235 & 134124 & 90815  & 105578 & 43926 & 20925 & 20403 \\
0.6  &  126248 & 30284 & 121780 & 146073 & 93686  & 68932 & 62787 & 19478 \\
0.55 &  131577 & 22127 & 123599 & 103778 & 88999  & 42183 & 28736 & 16607 \\
0.53 &  137302 & 24401 & 146940 & 84434  & 43674  & 64683 & 29323 & 15825 \\
0.5  &  128873 & 23648 & 99971  & 86445  & 26464  & 39693 & 23994 & 15355 \\
\end{tabular}
\end{ruledtabular}
\end{table}

\begin{table}[ht]
\caption{\label{tab:g2c03wt_sup3}
The measured gradient flow couplings $g_{\rm GF}^2$ with the chosen set of parameters:
LW evolved flow, clover definition of energy density, $c_t=0.3$ and $\tau_0=0.025\log(1+2*g_{GF}^2)$.
These are the parameters used in most of the work. The statistical errors are counted with the jackknife method.
}
\begin{ruledtabular}
\begin{tabular}{ccccccccc}
$\beta_L$ & $L=8$ & $L=10$ & $L=12$ & $L=16$ & $L=18$ & $L=20$ & $L=24$ & $L=30$ \\
\hline
8    & 0.56878(16) & 0.5639(5)  & 0.56447(17) & 0.5642(8)  & 0.5660(14) & 0.5660(9)  & 0.5674(5)  & 0.5687(19) \\ 
6    & 0.77786(18) & 0.7718(5)  & 0.7718(3)   & 0.7754(4)  & 0.7760(5)  & 0.7776(6)  & 0.7808(10) & 0.792(3)   \\ 
4    & 1.1816(3)   & 1.1736(11) & 1.1781(5)   & 1.1913(6)  & 1.1960(9)  & 1.2012(15) & 1.212(2)   & 1.224(5)   \\ 
3    & 1.5426(6)   & 1.5383(14) & 1.5546(9)   & 1.5827(12) & 1.5933(16) & 1.607(2)   & 1.630(4)   & 1.653(7)   \\ 
2    & 2.1936(10)  & 2.213(3)   & 2.2591(18)  & 2.329(2)   & 2.357(2)   & 2.379(4)   & 2.423(5)   & 2.466(12)  \\ 
1.7  & 2.5286(15)  & 2.559(3)   & 2.6261(19)  & 2.726(2)   & 2.765(4)   & 2.803(6)   & 2.863(7)   & 2.951(16)  \\ 
1.5  & 2.8258(14)  & 2.881(4)   & 2.956(2)    & 3.083(3)   & 3.142(3)   & 3.184(8)   & 3.268(9)   & 3.368(17)  \\ 
1.3  & 3.2215(19)  & 3.295(5)   & 3.396(3)    & 3.563(4)   & 3.638(6)   & 3.684(9)   & 3.781(9)   & 3.93(2)    \\ 
1    & 4.163(3)    & 4.283(6)   & 4.460(4)    & 4.730(6)   & 4.846(9)   & 4.939(14)  & 5.11(2)    & 5.33(3)    \\ 
0.9  & 4.665(3)    & 4.822(9)   & 5.008(5)    & 5.317(8)   & 5.456(12)  & 5.577(15)  & 5.82(2)    & 6.07(4)    \\ 
0.8  & 5.383(7)    & 5.538(14)  & 5.755(5)    & 6.145(16)  & 6.302(14)  & 6.451(18)  & 6.70(2)    & 7.12(5)    \\ 
0.7  & 6.570(8)    & 6.69(2)    & 6.867(10)   & 7.314(19)  & 7.509(15)  & 7.65(2)    & 8.06(3)    & 8.32(6)    \\ 
0.6  & 9.06(2)     & 8.83(3)    & 8.876(14)   & 9.16(2)    & 9.34(2)    & 9.53(2)    & 9.88(3)    & 10.36(7)   \\ 
0.55 & 12.86(4)    & 11.86(7)   & 11.39(2)    & 10.94(2)   & 10.98(2)   & 11.09(6)   & 11.37(6)   & 11.77(10)  \\ 
0.53 & 16.10(6)    & 15.59(16)  & 13.92(5)    & 12.43(4)   & 12.24(5)   & 12.24(6)   & 12.19(7)   & 12.85(12)  \\ 
0.5  & 22.04(5)    & 24.6(2)    & 22.97(11)   & 17.16(7)   & 15.80(12)  & 14.87(13)  & 14.12(10)  & 14.30(14)  \\ 
\end{tabular}
\end{ruledtabular}
\end{table}

\begin{table}[ht]
\caption{\label{tab:g2c03_sup4}
The measured gradient flow couplings $g_{\rm GF}^2$ with otherwise same parameters 
as~\ref{tab:g2c03wt_sup3} but with $\tau_0=0$.
}
\begin{ruledtabular}
\begin{tabular}{ccccccccc}
$\beta_L$ & $L=8$ & $L=10$ & $L=12$ & $L=16$ & $L=18$ & $L=20$ & $L=24$ & $L=30$ \\
\hline
8 & 0.60287(17) & 0.5851(5) & 0.57911(17) & 0.5724(8) & 0.5724(14) & 0.5712(9) & 0.5710(5) & 0.5711(19) \\ 
6 & 0.83581(19) & 0.8078(6) & 0.7966(3) & 0.7893(4) & 0.7869(5) & 0.7865(6) & 0.7870(10) & 0.796(3) \\ 
4 & 1.2955(3) & 1.2443(12) & 1.2268(5) & 1.2189(6) & 1.2178(9) & 1.2190(15) & 1.225(2) & 1.232(6) \\ 
3 & 1.7145(6) & 1.6456(15) & 1.6290(10) & 1.6252(12) & 1.6271(16) & 1.635(2) & 1.650(4) & 1.665(7) \\ 
2 & 2.4813(11) & 2.396(3) & 2.3875(18) & 2.404(2) & 2.417(2) & 2.428(4) & 2.458(5) & 2.488(12) \\ 
1.7 & 2.8793(16) & 2.782(3) & 2.784(2) & 2.819(2) & 2.840(4) & 2.864(6) & 2.907(7) & 2.980(16) \\ 
1.5 & 3.2339(15) & 3.143(4) & 3.143(2) & 3.193(3) & 3.231(3) & 3.257(8) & 3.321(9) & 3.403(17) \\ 
1.3 & 3.707(2) & 3.610(5) & 3.621(3) & 3.696(4) & 3.746(7) & 3.773(9) & 3.845(9) & 3.97(2) \\ 
1 & 4.839(3) & 4.725(7) & 4.780(4) & 4.923(6) & 5.002(9) & 5.069(14) & 5.21(2) & 5.40(3) \\ 
0.9 & 5.441(3) & 5.332(10) & 5.377(5) & 5.540(8) & 5.638(12) & 5.729(15) & 5.93(2) & 6.15(4) \\ 
0.8 & 6.297(7) & 6.139(15) & 6.191(6) & 6.410(16) & 6.518(14) & 6.632(18) & 6.83(2) & 7.21(5) \\ 
0.7 & 7.687(9) & 7.42(2) & 7.396(10) & 7.635(19) & 7.772(15) & 7.87(2) & 8.23(4) & 8.43(6) \\ 
0.6 & 10.58(2) & 9.80(3) & 9.562(15) & 9.56(2) & 9.67(2) & 9.81(2) & 10.08(3) & 10.50(7) \\ 
0.55 & 14.91(4) & 13.14(7) & 12.26(2) & 11.42(2) & 11.37(2) & 11.41(6) & 11.60(6) & 11.92(10) \\ 
0.53 & 18.57(6) & 17.21(16) & 14.95(5) & 12.97(4) & 12.66(5) & 12.59(6) & 12.44(7) & 13.02(12) \\ 
0.5 & 25.25(6) & 27.0(2) & 24.57(11) & 17.85(7) & 16.32(13) & 15.28(13) & 14.39(10) & 14.49(14) \\ 
\end{tabular}
\end{ruledtabular}
\end{table}

\begin{table}[ht]
\caption{\label{tab:g2c035_sup5}
The measured gradient flow couplings $g_{\rm GF}^2$ with otherwise same parameters 
as~\ref{tab:g2c03_sup4} but with $c_t=0.35$.
}
\begin{ruledtabular}
\begin{tabular}{ccccccccc}
$\beta_L$ & $L=8$ & $L=10$ & $L=12$ & $L=16$ & $L=18$ & $L=20$ & $L=24$ & $L=30$ \\
\hline
8 & 0.5911(2) & 0.5808(6) & 0.5771(2) & 0.5727(10) & 0.5739(19) & 0.5734(12) & 0.5739(6) & 0.574(2) \\ 
6 & 0.8194(2) & 0.8030(7) & 0.7957(4) & 0.7930(5) & 0.7917(7) & 0.7918(8) & 0.7933(13) & 0.804(4) \\ 
4 & 1.2754(5) & 1.2453(16) & 1.2341(8) & 1.2330(8) & 1.2336(13) & 1.236(2) & 1.245(3) & 1.249(7) \\ 
3 & 1.7015(8) & 1.659(2) & 1.6533(14) & 1.6575(16) & 1.661(2) & 1.672(3) & 1.690(5) & 1.711(10) \\ 
2 & 2.5077(16) & 2.459(4) & 2.462(2) & 2.491(3) & 2.507(3) & 2.518(6) & 2.549(8) & 2.583(16) \\ 
1.7 & 2.939(2) & 2.881(5) & 2.897(2) & 2.945(3) & 2.969(6) & 3.000(10) & 3.045(10) & 3.12(2) \\ 
1.5 & 3.330(2) & 3.284(6) & 3.294(3) & 3.359(5) & 3.404(5) & 3.436(13) & 3.502(13) & 3.59(2) \\ 
1.3 & 3.863(3) & 3.809(8) & 3.833(5) & 3.926(7) & 3.985(11) & 4.008(13) & 4.083(14) & 4.23(3) \\ 
1 & 5.174(6) & 5.101(11) & 5.186(7) & 5.351(10) & 5.451(15) & 5.51(2) & 5.68(4) & 5.90(5) \\ 
0.9 & 5.906(6) & 5.855(17) & 5.903(10) & 6.090(14) & 6.21(2) & 6.30(2) & 6.54(3) & 6.80(8) \\ 
0.8 & 7.005(14) & 6.87(2) & 6.928(11) & 7.18(3) & 7.30(2) & 7.42(3) & 7.64(5) & 8.15(9) \\ 
0.7 & 8.946(17) & 8.64(4) & 8.55(2) & 8.83(3) & 8.96(3) & 9.05(4) & 9.47(7) & 9.62(10) \\ 
0.6 & 13.12(3) & 12.06(5) & 11.66(2) & 11.54(3) & 11.62(4) & 11.76(4) & 12.02(6) & 12.48(12) \\ 
0.55 & 19.72(7) & 17.03(12) & 15.68(4) & 14.25(4) & 14.09(5) & 14.05(10) & 14.23(11) & 14.52(18) \\ 
0.53 & 25.51(10) & 23.4(2) & 19.89(8) & 16.67(7) & 16.11(9) & 15.90(11) & 15.50(12) & 16.1(2) \\ 
0.5 & 36.17(9) & 38.9(3) & 35.18(18) & 24.61(13) & 22.0(2) & 20.2(2) & 18.66(18) & 18.5(2) \\ 
\end{tabular}
\end{ruledtabular}
\end{table}

\begin{table}[ht]
\caption{\label{tab:g2c04_sup6}
The measured gradient flow couplings $g_{\rm GF}^2$ with otherwise same parameters 
as~\ref{tab:g2c03_sup4} but with $c_t=0.4$.
}
\begin{ruledtabular}
\begin{tabular}{ccccccccc}
$\beta_L$ & $L=8$ & $L=10$ & $L=12$ & $L=16$ & $L=18$ & $L=20$ & $L=24$ & $L=30$ \\
\hline
8 & 0.5851(2) & 0.5793(8) & 0.5774(3) & 0.5745(13) & 0.576(2) & 0.5770(16) & 0.5779(8) & 0.578(3) \\ 
6 & 0.8132(2) & 0.8035(9) & 0.7992(5) & 0.7999(6) & 0.7993(8) & 0.7998(10) & 0.8020(17) & 0.814(5) \\ 
4 & 1.2769(6) & 1.258(2) & 1.2511(10) & 1.2546(10) & 1.2563(18) & 1.260(2) & 1.272(5) & 1.271(9) \\ 
3 & 1.7227(11) & 1.693(2) & 1.6944(18) & 1.703(2) & 1.708(2) & 1.721(4) & 1.742(7) & 1.770(13) \\ 
2 & 2.597(2) & 2.564(6) & 2.572(3) & 2.609(4) & 2.627(5) & 2.638(9) & 2.670(12) & 2.70(2) \\ 
1.7 & 3.082(3) & 3.037(7) & 3.058(4) & 3.115(5) & 3.142(8) & 3.180(14) & 3.227(15) & 3.30(3) \\ 
1.5 & 3.531(3) & 3.498(10) & 3.509(4) & 3.583(8) & 3.638(8) & 3.676(18) & 3.743(19) & 3.84(3) \\ 
1.3 & 4.156(5) & 4.107(12) & 4.133(7) & 4.241(10) & 4.312(17) & 4.327(19) & 4.40(2) & 4.59(4) \\ 
1 & 5.760(10) & 5.66(2) & 5.779(12) & 5.959(15) & 6.09(2) & 6.13(3) & 6.35(6) & 6.60(8) \\ 
0.9 & 6.709(11) & 6.66(3) & 6.686(18) & 6.89(2) & 7.04(4) & 7.12(4) & 7.42(6) & 7.74(13) \\ 
0.8 & 8.24(2) & 8.03(5) & 8.06(2) & 8.35(6) & 8.48(5) & 8.62(6) & 8.87(9) & 9.60(18) \\ 
0.7 & 11.21(3) & 10.70(8) & 10.46(4) & 10.80(7) & 10.89(6) & 10.95(8) & 11.40(13) & 11.46(18) \\ 
0.6 & 17.58(6) & 15.98(9) & 15.29(4) & 14.94(6) & 14.96(7) & 15.07(8) & 15.29(10) & 15.7(2) \\ 
0.55 & 27.66(12) & 23.5(2) & 21.42(8) & 18.99(8) & 18.66(9) & 18.48(18) & 18.64(18) & 18.8(3) \\ 
0.53 & 36.70(17) & 33.6(4) & 28.14(14) & 22.90(13) & 21.90(17) & 21.4(2) & 20.6(2) & 21.3(3) \\ 
0.5 & 53.35(15) & 57.8(5) & 52.5(3) & 35.9(2) & 31.7(3) & 28.7(4) & 25.8(3) & 25.3(4) \\ 
\end{tabular}
\end{ruledtabular}
\end{table}

\begin{table}[ht]
\caption{\label{tab:g2c045_sup7}
The measured gradient flow couplings $g_{\rm GF}^2$ with otherwise same parameters 
as~\ref{tab:g2c03_sup4} but with $c_t=0.45$.
}
\begin{ruledtabular}
\begin{tabular}{ccccccccc}
$\beta_L$ & $L=8$ & $L=10$ & $L=12$ & $L=16$ & $L=18$ & $L=20$ & $L=24$ & $L=30$ \\
\hline
8 & 0.5832(2) & 0.5800(10) & 0.5794(3) & 0.5775(16) & 0.580(3) & 0.581(2) & 0.5831(10) & 0.583(3) \\ 
6 & 0.8142(3) & 0.8079(11) & 0.8060(6) & 0.8095(8) & 0.8094(11) & 0.8100(13) & 0.813(2) & 0.825(7) \\ 
4 & 1.2935(8) & 1.280(2) & 1.2761(13) & 1.2827(14) & 1.285(2) & 1.290(3) & 1.307(6) & 1.299(11) \\ 
3 & 1.7686(15) & 1.744(3) & 1.750(2) & 1.761(2) & 1.768(3) & 1.782(5) & 1.807(9) & 1.843(17) \\ 
2 & 2.737(3) & 2.706(9) & 2.718(4) & 2.762(5) & 2.780(7) & 2.792(13) & 2.822(17) & 2.86(2) \\ 
1.7 & 3.297(4) & 3.249(10) & 3.272(5) & 3.335(7) & 3.364(11) & 3.41(2) & 3.45(2) & 3.53(4) \\ 
1.5 & 3.828(6) & 3.788(14) & 3.793(6) & 3.874(12) & 3.939(12) & 3.98(2) & 4.05(2) & 4.17(4) \\ 
1.3 & 4.587(8) & 4.513(19) & 4.533(9) & 4.657(15) & 4.74(2) & 4.74(2) & 4.82(3) & 5.05(6) \\ 
1 & 6.644(19) & 6.48(3) & 6.61(2) & 6.80(2) & 6.99(4) & 7.00(6) & 7.29(10) & 7.59(14) \\ 
0.9 & 7.95(2) & 7.87(5) & 7.82(3) & 8.04(4) & 8.25(9) & 8.30(7) & 8.70(12) & 9.1(2) \\ 
0.8 & 10.27(6) & 9.85(11) & 9.80(4) & 10.14(11) & 10.30(10) & 10.43(12) & 10.72(17) & 11.8(3) \\ 
0.7 & 15.28(7) & 14.20(17) & 13.68(8) & 14.10(15) & 14.09(13) & 14.07(18) & 14.5(2) & 14.4(3) \\ 
0.6 & 25.73(10) & 23.15(15) & 21.96(8) & 21.14(10) & 21.04(12) & 21.07(14) & 21.17(16) & 21.6(3) \\ 
0.55 & 41.4(2) & 34.9(3) & 31.59(14) & 27.40(14) & 26.75(16) & 26.3(3) & 26.4(3) & 26.4(5) \\ 
0.53 & 55.6(2) & 51.1(7) & 42.5(2) & 33.8(2) & 32.0(3) & 31.2(3) & 29.6(3) & 30.3(6) \\ 
0.5 & 81.7(2) & 89.6(9) & 82.0(4) & 55.5(3) & 48.5(6) & 43.4(7) & 38.4(5) & 37.1(7) \\ 
\end{tabular}
\end{ruledtabular}
\end{table}

\begin{table}[ht]
\caption{\label{tab:g2c03W_sup8}
The measured gradient flow couplings $g_{\rm GF}^2$ with otherwise same parameters 
as~\ref{tab:g2c03_sup4} but with wilson flow (W).
}
\begin{ruledtabular}
\begin{tabular}{ccccccccc}
$\beta_L$ & $L=8$ & $L=10$ & $L=12$ & $L=16$ & $L=18$ & $L=20$ & $L=24$ & $L=30$ \\
\hline
8 & 0.7777(2) & 0.6961(5) & 0.65497(18) & 0.6141(8) & 0.6053(14) & 0.5977(9) & 0.5894(5) & 0.5828(19) \\ 
6 & 1.0912(2) & 0.9682(6) & 0.9055(3) & 0.8492(4) & 0.8340(5) & 0.8245(6) & 0.8135(10) & 0.813(3) \\ 
4 & 1.7257(4) & 1.5102(13) & 1.4069(6) & 1.3182(6) & 1.2960(9) & 1.2823(15) & 1.269(2) & 1.261(6) \\ 
3 & 2.3150(7) & 2.0152(16) & 1.8797(10) & 1.7643(12) & 1.7370(16) & 1.724(2) & 1.712(4) & 1.706(7) \\ 
2 & 3.4037(13) & 2.966(3) & 2.7779(19) & 2.623(2) & 2.591(2) & 2.570(4) & 2.558(6) & 2.554(12) \\ 
1.7 & 3.9705(19) & 3.458(3) & 3.250(2) & 3.083(2) & 3.050(4) & 3.036(7) & 3.029(7) & 3.061(16) \\ 
1.5 & 4.4776(17) & 3.919(4) & 3.677(2) & 3.497(3) & 3.474(4) & 3.456(9) & 3.463(9) & 3.497(17) \\ 
1.3 & 5.155(2) & 4.517(6) & 4.249(3) & 4.056(5) & 4.034(7) & 4.009(9) & 4.014(9) & 4.09(2) \\ 
1 & 6.778(4) & 5.953(7) & 5.634(4) & 5.418(6) & 5.401(9) & 5.398(14) & 5.45(2) & 5.56(3) \\ 
0.9 & 7.633(4) & 6.724(11) & 6.350(5) & 6.105(8) & 6.094(13) & 6.106(15) & 6.20(2) & 6.33(4) \\ 
0.8 & 8.827(8) & 7.752(15) & 7.318(6) & 7.068(16) & 7.050(14) & 7.072(19) & 7.15(2) & 7.42(5) \\ 
0.7 & 10.691(9) & 9.34(2) & 8.735(10) & 8.413(19) & 8.401(15) & 8.39(2) & 8.60(4) & 8.66(6) \\ 
0.6 & 14.40(2) & 12.24(3) & 11.227(16) & 10.50(2) & 10.42(2) & 10.42(2) & 10.48(3) & 10.69(7) \\ 
0.55 & 19.60(5) & 16.15(8) & 14.24(3) & 12.46(2) & 12.18(2) & 12.03(6) & 11.98(6) & 12.01(10) \\ 
0.53 & 23.82(7) & 20.77(18) & 17.17(5) & 14.05(4) & 13.48(5) & 13.20(6) & 12.77(7) & 12.97(12) \\ 
0.5 & 31.43(6) & 31.9(2) & 27.60(12) & 19.01(7) & 17.10(13) & 15.82(13) & 14.61(10) & 14.25(13) \\
\end{tabular}
\end{ruledtabular}
\end{table}

\begin{table}[ht]
\caption{\label{tab:g2c03P_sup9}
The measured gradient flow couplings $g_{\rm GF}^2$ with otherwise same parameters as~\ref{tab:g2c03_sup4} 
but with Plaquette measurement of energy density.
}
\begin{ruledtabular}
\begin{tabular}{ccccccccc}
$\beta_L$ & $L=8$ & $L=10$ & $L=12$ & $L=16$ & $L=18$ & $L=20$ & $L=24$ & $L=30$ \\
\hline
8 & 0.8059(2) & 0.6984(6) & 0.6528(2) & 0.6117(8) & 0.6030(15) & 0.5955(9) & 0.5878(5) & 0.581(2) \\ 
6 & 1.1209(2) & 0.9649(6) & 0.8986(3) & 0.8434(4) & 0.8289(5) & 0.8204(6) & 0.8102(10) & 0.810(3) \\ 
4 & 1.7448(4) & 1.4880(14) & 1.3845(6) & 1.3027(6) & 1.2833(9) & 1.2714(15) & 1.260(2) & 1.255(6) \\ 
3 & 2.3125(7) & 1.9679(17) & 1.8379(10) & 1.7368(12) & 1.7149(16) & 1.705(2) & 1.699(4) & 1.696(7) \\ 
2 & 3.3486(13) & 2.865(3) & 2.6927(19) & 2.569(2) & 2.546(2) & 2.533(4) & 2.531(6) & 2.534(12) \\ 
1.7 & 3.885(2) & 3.328(3) & 3.140(2) & 3.012(2) & 2.992(4) & 2.988(7) & 2.994(7) & 3.037(16) \\ 
1.5 & 4.365(2) & 3.758(5) & 3.545(2) & 3.412(3) & 3.403(4) & 3.397(9) & 3.420(9) & 3.468(17) \\ 
1.3 & 5.008(2) & 4.316(6) & 4.084(3) & 3.949(5) & 3.947(7) & 3.935(9) & 3.959(9) & 4.05(2) \\ 
1 & 6.549(4) & 5.653(7) & 5.392(4) & 5.262(6) & 5.272(9) & 5.287(14) & 5.36(2) & 5.50(3) \\ 
0.9 & 7.376(4) & 6.378(11) & 6.066(6) & 5.921(9) & 5.943(13) & 5.977(15) & 6.11(2) & 6.26(5) \\ 
0.8 & 8.555(9) & 7.348(16) & 6.986(6) & 6.851(16) & 6.868(14) & 6.918(19) & 7.04(3) & 7.35(5) \\ 
0.7 & 10.492(11) & 8.89(2) & 8.351(10) & 8.16(2) & 8.192(15) & 8.22(2) & 8.48(4) & 8.59(6) \\ 
0.6 & 14.78(3) & 11.85(4) & 10.850(16) & 10.25(2) & 10.21(2) & 10.25(2) & 10.41(3) & 10.73(8) \\ 
0.55 & 21.61(7) & 16.18(9) & 14.04(3) & 12.29(3) & 12.04(3) & 11.95(6) & 12.00(6) & 12.22(10) \\ 
0.53 & 27.60(10) & 21.5(2) & 17.26(6) & 14.00(4) & 13.43(5) & 13.22(7) & 12.89(7) & 13.37(13) \\ 
0.5 & 39.08(10) & 35.2(3) & 29.15(14) & 19.43(8) & 17.43(14) & 16.12(13) & 14.96(11) & 14.91(14) \\ 
\end{tabular}
\end{ruledtabular}
\end{table}

\begin{table}[ht]
\caption{\label{tab:g2c03PW_sup10}
The measured gradient flow couplings $g_{\rm GF}^2$ with otherwise same parameters as~\ref{tab:g2c03_sup4} 
but with Plaquette measurement of energy density and with wilson flow (W).
}
\begin{ruledtabular}
\begin{tabular}{ccccccccc}
$\beta_L$ & $L=8$ & $L=10$ & $L=12$ & $L=16$ & $L=18$ & $L=20$ & $L=24$ & $L=30$ \\
\hline
8 & 1.1442(2) & 0.8610(6) & 0.7497(2) & 0.6592(8) & 0.6393(15) & 0.6242(9) & 0.6073(5) & 0.593(2) \\ 
6 & 1.6195(2) & 1.2003(7) & 1.0377(3) & 0.9115(4) & 0.8809(5) & 0.8616(6) & 0.8381(10) & 0.828(3) \\ 
4 & 2.5970(5) & 1.8793(15) & 1.6142(6) & 1.4154(6) & 1.3696(10) & 1.3400(15) & 1.307(2) & 1.285(6) \\ 
3 & 3.5106(8) & 2.5118(18) & 2.1574(11) & 1.8946(13) & 1.8363(17) & 1.802(2) & 1.765(4) & 1.738(7) \\ 
2 & 5.2024(16) & 3.705(4) & 3.190(2) & 2.818(2) & 2.738(2) & 2.687(4) & 2.637(6) & 2.602(12) \\ 
1.7 & 6.086(2) & 4.325(4) & 3.733(2) & 3.311(2) & 3.223(4) & 3.174(7) & 3.123(7) & 3.120(16) \\ 
1.5 & 6.882(2) & 4.902(5) & 4.227(2) & 3.757(3) & 3.671(4) & 3.612(9) & 3.570(9) & 3.565(17) \\ 
1.3 & 7.954(3) & 5.657(6) & 4.885(3) & 4.357(5) & 4.265(7) & 4.190(9) & 4.137(9) & 4.16(2) \\ 
1 & 10.554(5) & 7.476(8) & 6.484(5) & 5.824(7) & 5.713(9) & 5.644(14) & 5.61(2) & 5.67(3) \\ 
0.9 & 11.947(6) & 8.454(12) & 7.313(6) & 6.563(9) & 6.446(13) & 6.386(15) & 6.39(2) & 6.45(5) \\ 
0.8 & 13.914(10) & 9.771(18) & 8.436(6) & 7.600(17) & 7.456(15) & 7.396(19) & 7.38(3) & 7.57(5) \\ 
0.7 & 17.038(14) & 11.82(2) & 10.083(11) & 9.05(2) & 8.889(15) & 8.78(2) & 8.87(4) & 8.83(6) \\ 
0.6 & 23.68(4) & 15.69(4) & 13.037(17) & 11.32(2) & 11.04(2) & 10.91(2) & 10.83(3) & 10.92(8) \\ 
0.55 & 33.67(10) & 21.21(11) & 16.71(3) & 13.48(3) & 12.94(3) & 12.62(6) & 12.39(6) & 12.27(10) \\ 
0.53 & 42.17(14) & 27.9(2) & 20.37(6) & 15.25(4) & 14.34(5) & 13.87(7) & 13.22(7) & 13.26(12) \\ 
0.5 & 58.45(14) & 45.2(4) & 33.94(16) & 20.82(9) & 18.30(14) & 16.70(13) & 15.14(11) & 14.59(14) \\ 
\end{tabular}
\end{ruledtabular}
\end{table}

\begin{table}[ht]
\caption{\label{tab:zp_sup11}
The measured bare values of $Z_P$ for each lattice size $L$ and $\beta_L$. 
The step scaling mass anomalous dimension is computed from these using the steps given in the main text.
}
\begin{ruledtabular}
\begin{tabular}{ccccccccc}
$\beta_L$ & $L=8$ & $L=10$ & $L=12$ & $L=16$ & $L=18$ & $L=20$ & $L=24$ & $L=30$ \\
\hline
8    & 0.97103(6) & 0.9670(2)  & 0.96430(10) & 0.9600(3)   & 0.9578(6)  & 0.9568(3)  & 0.9534(2)  & 0.9499(6)  \\ 
6    & 0.95990(8) & 0.9545(2)  & 0.95067(14) & 0.94396(12) & 0.9413(2)  & 0.9405(3)  & 0.9358(4)  & 0.9298(12) \\ 
4    & 0.991(3)   & 0.9279(4)  & 0.9646(2)   & 0.9130(2)   & 0.9097(3)  & 0.9055(6)  & 0.9009(11) & 0.893(2)   \\ 
3    & 0.9135(2)  & 0.9023(5)  & 0.8953(3)   & 0.8835(4)   & 0.8794(4)  & 0.8734(9)  & 0.8684(12) & 0.860(2)   \\ 
2    & 0.8749(3)  & 0.8615(9)  & 0.8481(5)   & 0.8328(6)   & 0.8244(8)  & 0.8176(13) & 0.8124(16) & 0.790(3)   \\ 
1.7  & 0.8557(4)  & 0.8408(11) & 0.8256(6)   & 0.8080(8)   & 0.7988(8)  & 0.7920(16) & 0.7833(18) & 0.765(3)   \\ 
1.5  & 0.8407(6)  & 0.8204(13) & 0.8068(6)   & 0.7859(8)   & 0.7780(9)  & 0.7696(18) & 0.754(2)   & 0.753(3)   \\ 
1.3  & 0.8219(8)  & 0.7979(16) & 0.7827(7)   & 0.7601(7)   & 0.7506(12) & 0.746(2)   & 0.7316(19) & 0.708(4)   \\ 
1    & 0.7734(8)  & 0.749(2)   & 0.7304(10)  & 0.7016(17)  & 0.6902(18) & 0.677(2)   & 0.664(4)   & 0.650(6)   \\ 
0.9  & 0.7468(10) & 0.717(2)   & 0.7010(10)  & 0.6733(18)  & 0.660(2)   & 0.646(2)   & 0.632(2)   & 0.615(6)   \\ 
0.8  & 0.703(2)   & 0.680(3)   & 0.6603(14)  & 0.6344(18)  & 0.616(2)   & 0.609(3)   & 0.595(4)   & 0.576(7)   \\ 
0.7  & 0.618(2)   & 0.597(5)   & 0.588(2)    & 0.548(3)    & 0.546(3)   & 0.539(5)   & 0.515(7)   & 0.498(10)  \\ 
0.6  & 0.411(2)   & 0.401(6)   & 0.376(4)    & 0.346(3)    & 0.338(9)   & 0.328(10)  & 0.293(7)   & 0.283(15)  \\ 
0.55 & 0.3450(13) & 0.323(4)   & 0.3068(16)  & 0.275(2)    & 0.267(2)   & 0.252(3)   & 0.253(5)   & 0.239(8)   \\ 
0.53 & 0.3193(10) & 0.299(3)   & 0.2842(11)  & 0.2619(16)  & 0.2514(19) & 0.249(2)   & 0.228(2)   & 0.227(7)   \\ 
0.5  & 0.2803(7)  & 0.254(2)   & 0.2425(7)   & 0.2356(9)   & 0.2345(18) & 0.229(2)   & 0.222(2)   & 0.209(4)   \\ 
\end{tabular}
\end{ruledtabular}
\end{table}%

\begin{table}[ht]
\caption{\label{tab:irfptable}
Location of the IRFP for different discretizations: LW=Lüscher-Weisz, W=Wilson, $\tau_0=\tau_0-\mathrm{correction}$, C=Clover
and P=Plaquette. 
}
\begin{ruledtabular}
\begin{tabular}{cccccc}
$c_t$ & $LWC\tau_0$ & $LWC$ & $LCP$ & $WC$ & $WP$  \\
\hline
0.3   & 14.5(4) & 14.1(3) & 14.3(2) & 13.5(2) & 14.0(2)  \\
0.35  & 17.1(5) & 17.1(2) & 17.5(4) & 16.1(3) & 16.4(3)  \\
0.4   & 22.2(6) & 22.3(6) & 22.9(6) & 20.2(5) & 20.5(5)  \\
0.45  & 31(1)   & 31.1(9) & 32(1)   & 27(14)  & 26(15)   \\
\end{tabular}
\end{ruledtabular}
\end{table}%

\begin{table}[ht]
\caption{\label{tab:irfpstar}
Value of $\gamma_g^\ast$ measured from slope of the $\beta$-function 
for different discretizations: LW=Lüscher-Weisz, W=Wilson, $\tau_0=\tau_0-\mathrm{correction}$, C=Clover
and P=Plaquette. 
}
\begin{ruledtabular}
\begin{tabular}{cccccc}
$c_t$ & $LWC\tau_0$ & $LWC$ & $LCP$ & $WC$ & $WP$  \\
\hline
0.3   & 0.648(97) & 0.68(9)   & 0.74(10) & 0.8(1)   & 0.77(10)   \\
0.35  & 0.71(12)  & 0.699(85) & 0.69(9)  & 0.76(12) & 0.70(12)   \\
0.4   & 0.73(10)  & 0.74(10)  & 0.74(10) & 0.69(14) & 0.59(16)   \\
0.45  & 0.75(12)  & 0.75(11)  & 0.74(11) & 0.51(39) & 0.40(28)   \\
\end{tabular}
\end{ruledtabular}
\end{table}

\begin{table}[ht]
\caption{
Combined $\chi^2/$d.o.f of the beta interpolations in use \eqref{eq:betafitfun} 
at different $c_t$ and available discretization options.
}
\begin{ruledtabular}
\begin{tabular}{cccccc}
$c_t$ & $LWC\tau_0$ & $LWC$ & $LCP$ & $WC$ & $WP$  \\
\hline 
0.3   & 1.29  & 1.32  & 1.57 & 1.56 & 2.51  \\
0.35  & 1.09  & 1.11  & 1.21 & 1.19 & 1.51  \\
0.4   & 1.12  & 1.12  & 1.17 & 1.16 & 1.29  \\
0.45  & 1.43  & 1.41  & 1.45 & 1.41 & 1.47  \\
\end{tabular}
\end{ruledtabular}
\label{tab:combchi}
\end{table}
\end{document}